\documentclass[a4paper,11pt]{article}
\pdfoutput=1 
\usepackage{jheppub} 

\usepackage[T1]{fontenc} 

\title{Light-cone distribution amplitudes of the nucleon and $\Delta$
  baryon\footnote{On August 25, 2021, our respected and beloved
    colleague and friend  Maxim Vladimirovich Polyakov passed
    away. The present work has been done under his coaching. We
    dedicate it to the memory of our friend Maxim Vladimirovich
    Polyakov.}}    
\usepackage{graphicx}  
\usepackage{dcolumn}  
\usepackage{bm}          
\usepackage{slashed}
\usepackage{cancel}
\usepackage{float}
\usepackage{mathtools}
\usepackage{amsbsy}
\usepackage{amstext}

\usepackage[utf8]{inputenc} 
\usepackage{booktabs}
\usepackage[normalem]{ulem} 
\usepackage[dvipsnames]{xcolor} 
\usepackage{tabularx}
\usepackage{enumitem}  
\usepackage{array} 
\usepackage{slashed}
\usepackage{tikz}
\usepackage{float}
\usepackage{multirow}
\usepackage{subfig}
\renewcommand\sout{\bgroup \color{red} \ULdepth=-.5ex \ULset}

\makeatletter

\preprint{INHA-NTG-09/2021}

\author[a]{June-Young Kim,}
\author[b,c]{Hyun-Chul Kim}
\author[,a,d]{Maxim V. Polyakov\footnote{Deceased.}}

\affiliation[a]{Institut f\"ur Theoretische Physik II, Ruhr-Universit\"at
  Bochum, D-44780 Bochum, Germany}
\affiliation[b]{Department of Physics, Inha University, Incheon 22212,
 Republic of Korea}
\affiliation[c]{School of Physics, Korea Institute for Advanced Study 
   (KIAS), Seoul 02455, Republic of Korea}
\affiliation[d]{Petersburg Nuclear Physics Institute, Gatchina, 188300,
  St. Petersburg, Russia}

\emailAdd{Jun-Young.Kim@ruhr-uni-bochum.de}
\emailAdd{hchkim@inha.ac.kr}

\abstract{ We investigate the light-cone wave functions and leading-twist
  distribution amplitudes for the nucleon and $\Delta$ baryon within the
  framework of the  chiral quark-soliton model. The baryon wave function
  consists of the valence quark  and vacuum wave functions. The vacuum 
  wave functions generate all possible higher Fock states by expanding them.
  We find that it is essential to consider the five-quark component and relativistic
  corrections to evaluate the distribution amplitudes of the nucleon and
  $\Delta$ isobar. Having taken into account them,
  we derive the distribution amplitudes. The results are in good agreement with
  the lattice data.}

\begin{document} 

\maketitle
\flushbottom
\section{Introduction\label{sec:1}}
Understanding the quark-gluon structure of the nucleon has been one of
the most important issues in hadronic physics. The nucleon light-cone
wave function (LCWF) provides a systematic way of examining it, since
it is decomposed unequivocally by the Fock states of quark-gluon
degrees of freedom on the light cone~\cite{Brodsky:1997de,
  Lepage:1980fj, Chernyak:1983ej}. This LCWF is derived 
by  projecting the nucleon state on the Fock basis that consists of
various partons, i.e., quark ($Q$),  antiquark ($\bar{Q}$) and gluon
($g$). The LCWF is considered as a basic building block in
high-energy reactions. The $3Q$ Fock component of the LCWF arises as
the most essential one~\cite{Dziembowski:1987es, Ji:2002xn} for the
nucleon, whereas higher Fock states come next~\cite{Ji:2003yj}. 
The LCWF in the context of quantum chromodynamics~(QCD) is deeply
rooted in the factorization theorem. It represents the soft part that
is crucial in describing any high-energy processes. In general, the
nonperturbative  observables such as generalized parton
distributions~(GPDs), the parton distributions functions~(PDFs), form
factors~(FF), and fragmentation functions are formulated in  terms of
the hadronic matrix element of the non-local QCD operators, and all
these quantities can be in principle constructed based on the
LCWFs. For example, the GPDs~\cite{Boffi:2002yy, Boffi:2003yj,
  Pasquini:2005dk, Pasquini:2007xz, Lorce:2011dv}, the
PDFs~\cite{Lepage:1980fj, Bolz:1996sw, Diehl:1998kh, Braun:2011aw,
  Pasquini:2018oyz} and the  transverse momentum-dependent parton
distributions (TMDs)~\cite{Pasquini:2008ax, Ji:2002xn, Lorce:2011dv}
are accessed by the overlap representations of the LCWFs. 
Essential information on the LCWFs is contained in the distribution
amplitudes (DAs), which encode how the quarks inside a hadron carry
the fraction of the longitudinal momentum of the hadron. In this work,
we want to investigate the DAs of the nucleon and $\Delta$ isobar. 

The DAs have been intensively developed so that one can investigate
the elastic or transition form factors at large momentum transfer for
a meson~\cite{Lepage:1979zb, Efremov:1979qk, Farrar:1979aw} and  a
baryon~\cite{Duncan:1979hi, Brodsky:1980sx, Lepage:1980fj,
  Chernyak:1983ej}. They are formally defined as a vacuum-to-baryon
matrix element of a nonlocal operator~\cite{Chernyak:1983ej}. Since the
contributions from the higher Fock components are power-suppressed,
they are less important at larger momentum transfer. The form factors
can be approximated to the convolution of the hard kernel of an
exclusive process with the involved DAs from the leading Fock
state. We want to mention that in the valence quark configuration the
leading-twist DAs are relevant to the LCWFs with the orbital angular
momentum $L_{z}=0$, whereas the higher-twist ones are connected to
those with both $L_{z}=0$ and $L_{z}\neq0$.   

There has been a great number of theoretical works on the DAs. The
dynamical properties of the nucleon DAs were investigated by employing
the QCD sum rules~\cite{Chernyak:1984bm,Gari:1986dr,  King:1986wi,
  Carlson:1986zs, Chernyak:1987nt, Chernyak:1987nu, Bergmann:1993eu,
  Braun:2001tj} 
(see also reviews~\cite{Chernyak:1983ej, Stefanis:1997zyh}).
The effects of explicit SU(3) symmetry breaking~\cite{Chernyak:1987nu,
  Wein:2015oqa} on the hyperon DAs and those form factors were
evaluated in Refs.~\cite{Liu:2008yg, Liu:2008zi, Liu:2009uc,
  Chernyak:1987nu}. The formalism for the DAs of the baryon decuplet
was first explored in the QCD sum rules and the perturbative QCD
predictions of the cross sections $\gamma \gamma \to B\bar{B}$ were
obtained in Ref.~\cite{Farrar:1988vz} (see also
Refs.~\cite{Braun:1999te, Pire:2011xv, Pire:2021hbl}).  The
higher-twist DAs were systematically studied for the nucleon in
Ref.~\cite{Braun:2000kw} and were associated  with the electromagnetic
form factors 
in Refs.~\cite{Braun:2001tj, Braun:2006hz, Aliev:2013jda,
  Anikin:2013aka}. Lattice QCD also provides invaluable information on
the baryon octet DAs~\cite{QCDSF:2008qtn, QCDSF:2008zfe,
  Gockeler:2008xv, Braun:2014wpa, Bali:2015ykx, RQCD:2019hps}.
The proton and Roper DAs were studied in the framework of
Dyson-Schwinger equations\cite{Mezrag:2017znp, Mezrag:2018hkk}.
Moreover, various phenomenological
approaches~\cite{Dziembowski:1987zq, Stefanis:1992pi, Stefanis:1992nw,
  Bergmann:1993rz, Bolz:1994fs, Bolz:1996sw, Wong:1999hc,
  Pasquini:2009ki} have been developed, the results from lattice QCD
and QCD sum rule being taken into account. 

In this paper, we aim at providing information on the nucleon and
$\Delta$ baryon DAs, and the corresponding normalization constants to
the leading-twist accuracy by employing the chiral quark-soliton model
($\chi$QSM). The $\chi$QSM, motivated by large $N_{c}$
QCD~\cite{Witten:1979kh, Witten:1983tw}, is a pion mean-field
theory. The presence of the $N_{c}$ valence quarks creates the pion
mean  field, which comes from the vacuum polarization of the Dirac
continuum. The $N_c$ valence quarks are then self-consistently
influenced by this pion mean field.  As a result, a baryon appears as
a chiral soliton that consists of the $N_{c}$ valence quarks bound by
the pion mean field. The $\chi$QSM has been
developed~\cite{Diakonov:1987ty, Wakamatsu:1990ud, Christov:1995vm},
based on the QCD instanton vacuum~\cite{Diakonov:1983hh,
  Diakonov:2002fq}. It respects important properties of the low-energy
QCD such as chiral symmetry and its spontaneous breaking. The
$\chi$QSM was successful in describing numerous low-energy properties
of the baryon~\cite{Christov:1995vm, Goeke:2007fp}. In addition, the
\emph{time-dependent} mean fields describing the moving soliton with
velocity $v\to1$ are obtained from the stationary mean fields by the
Lorentz boost and are applied to the nucleon
PDFs~\cite{Diakonov:1996sr, Diakonov:1997vc, Wakamatsu:1997en,
  Wakamatsu:1998rx, Schweitzer:2001sr, Son:2019ghf} and
GPDs~\cite{Petrov:1998kf, Goeke:2001tz}. 
These PDFs and GPDs from the $\chi$QSM comply with all general
theorems and sum rules. In fact, the nucleon LCWF was first derived in
Ref.~\cite{Petrov:2002jr} in this framework and its large-$N_{c}$
features were studied in Ref.~\cite{Pobylitsa:2004xj,
  Pobylitsa:2005rq}. Based on the baryon LCWFs from the model, various
observables~\cite{Diakonov:2004as, Diakonov:2005ib, 
  Lorce:2006nq, Lorce:2007fa, Lorce:2007as, Lorce:2011dv, Lorce:2011kd,
  Lorce:2011ni} were scrutinized by means of the overlap integrals. In
this context, it is of great importance to examine the nucleon and
$\Delta$ baryon DAs,  which yield yet another essential information on
the internal structures of the nucleon and $\Delta$ baryon.  

We sketch the present work as follows: In Sec.~\ref{sec:2} we briefly
review the definition of the nucleon DAs from the vacuum-to-nucleon
matrix element of the trilocal QCD operator and discuss the symmetry
properties of the nucleon DAs. In Sec.~\ref{sec:3}, the $\Delta$
baryon DAs is also derived in terms of the  vacuum-to-$\Delta$ matrix
element, and their symmetry properties are examined. The formalism for
the LCWFs and DAs of the nucleon and $\Delta$ baryon is constructed
within the framework of the $\chi$QSM in Sec.~\ref{sec:4}. 
The numerical results are presented and discussed in
Sec.~\ref{sec:5}. The final Sec.~\ref{sec:6} is devoted to summary and
conclusions. 

\section{Nucleon distribution amplitudes \label{sec:2}}

We start with a brief review on the general decomposition of the
vacuum-to-nucleon matrix element of the trilocal quark field
operators. It involves $24$ different invariant functions, which was
first investigated in Ref.~\cite{Braun:2000kw}. The matrix element of
the three-quark operator on the light cone reads 
\begin{align}
\langle 0 | \epsilon^{ijk} u^{i'}_{\alpha}(a_{1}n) [a_1,a_0]_{i'i}
  u^{j'}_{\beta}(a_{2}n) [a_2,a_0]_{j'j} d^{k'}_{\gamma}(a_{3}n)
  [a_3,a_0]_{k'k} |  N^{+} (p_{N},\lambda) \rangle, 
\label{eq:3qoperator}
\end{align}
where $|  N^{+} (p_{N},\lambda) \rangle$ stands for the
proton state with its momentum $p_{N}$
$(p^{2}_{N}=M^{2}_{N})$ and helicity $\lambda$, respectively. The
Greek letters $\alpha, \beta, \gamma$ denote the Dirac indices,
whereas the Latin ones $i,j,k$ designate the color indices. $n$
represents an arbitrary light-like vector ($n^{2}=0$) and $a_{i}$ is
specified as the quark separation between quarks. 
In order to be a gauge-invariant matrix element, the gauge connection
$[z_i,z_0]$ should be introduced:  
\begin{align}
[a_i,a_0] = P \exp[ig(a_{i}-a_{0})\int^{1}_{0} dt n_{\mu}A^{\mu}(n[ta_{i}+(1-t)a_{0}])],
\label{eq:gaugefactor}
\end{align} 
where $P$ stands for the path-ordering. Since we choose the light-cone
gauge $n\cdot A = 0$, however, the gauge connection becomes the
identity.  Considering the explicit Lorentz-covariant parametrization
with parity symmetry and the spin of the baryon taken into account, we
are able to sort out the 24 different invariant functions, which are
reduced to the three leading-twist ones:  
\begin{align}
  &4\langle 0 | \epsilon^{ijk} u^{i}_{\alpha}(a_{1}n)
    u^{j}_{\beta}(a_{2}n) 
    d^{k}_{\gamma}(a_{3}n)  |  N^{+} (p_{N},\lambda)
    \rangle \cr 
  &= f_{N} \bigg{[}(\slashed{p}_{N}C)_{\alpha \beta}
    (\gamma_{5}N)_{\gamma}V_{N} 
    (a_{i}n\cdot p_{N}) +
    (\slashed{p}_{N}\gamma_{5}C)_{\alpha\beta}N_{\gamma}
    A_N(a_{i}n\cdot p_{N}) \cr 
  & \hspace{1.2cm}]
    +(i\sigma_{\mu\nu}p^{\nu}_{N}C)_{\alpha\beta}(
    \gamma^{\mu}\gamma_{5}N)_{\gamma}T_N(a_{i}n\cdot p_{N}) \bigg{]},  
\label{eq:matrixoctet}
\end{align}
where $C$ is the charge conjugation matrix and
$\sigma_{\mu\nu}=i[\gamma_{\mu},\gamma_{\nu}]/2$. $N$ is the nucleon
spinor that satisfies the Dirac equation, i.e., $\slashed{p}_{N}N =
M_{N} N$, and is normalized to be $\bar{N}N=2M_{N}$. $f_{N}$ is known
to be the nucleon decay constant or the DA normalization constant,
which is identical to the value of the matrix element at the origin,
given in Eq.~\eqref{eq:matrixoctet}. The  dimensionless functions
$V_{N}, A_{N}$ and $T_{N}$ are  the Lorentz scalars that depend on
$p_{N}\cdot a_{i}$, and are normalized to be 
\begin{align}
V_{N}(0)=T_{N}(0)=1, \ \ \ A_{N}(0)=0.
\end{align} 
In the leading-twist accuracy, the nucleon momentum $p^{\mu}_{N}$ can be approximated to the light-cone vector.  The scalar functions $F=V_{N}$ ($A_{N},T_{N}$) can be defined in the momentum space by the Fourier transforms:
\begin{align}
  &F({x}_{i}) \delta(1-\sum^{3}_{l=1} x_{l}) = (n\cdot p_{N})^{3} \int \prod^{3}_{j=1} \frac{d a_{j}}{(2\pi)^{3}} F(a_{i}n\cdot p_{N}) \exp\bigg{[}i x_{k} a_{k}(n\cdot p_{N})\bigg{]}, \cr
&F(a_{i}n\cdot p_{N}) = \int \prod^{3}_{j=1} d x_{j} \delta(1-\sum^{3}_{l=1} x_{l}) F({x}_{i}) \exp\bigg{[}-i x_{k} a_{k}(n\cdot p_{N})\bigg{]},
 \label{eq:Fourier}
\end{align}
where the variables $x_i$ denote the fractions of the baryon longitudinal momentum, carried by the partons on the light cone. They satisfy $\sum^{3}_{i=1} x_{i}=1$ and
$ 0\leq x_{i} \leq 1$ by momentum conservation. The DAs $V_N$, $A_N$, and $T_N(x_{i})$ are defined at a certain scale $\mu$.
For simplicity, we suppress it. 

Not all the three scalar functions are independent. Exchanging the first two quarks
and considering the properties of the transpose of the Dirac matrices in Eq.~\eqref{eq:matrixoctet}, we find that $V_{N},T_{N}$ are symmetric whereas $A_{N}$ becomes antisymmetric: 
\begin{align}
&V_{N}(x_1,x_2,x_3)=V_{N}(x_2,x_1,x_3), \ \ \ A_{N}(x_1,x_2,x_3)=-A_{N}(x_2,x_1,x_3), \cr
&T_{N}(x_1,x_2,x_3)=T_{N}(x_2,x_1,x_3).
\label{eq:sym_N}
\end{align}
In addition, we impose the requirement on the three coupled quarks to give an isospin
1/2 state for the nucleon, which yields the following relation 
\begin{align}
2T_{N}(x_1,x_2,x_3)=[V-A]_{N}(x_1,x_3,x_2) + [V-A]_{N}(x_2,x_3,x_1).
\end{align}
Thus, we are able to express the nucleon DAs in terms of the single scalar function
defined as $\varphi_{N}(x_1,x_2,x_3):= [V-A]_{N}(x_1,x_2,x_3)$.
In flavor SU(3) symmetry, the given relation is valid for the baryon octet except for 
the isosinglet $\Lambda$ baryon. Therefore, all the hyperon DAs can be obtained by
considering flavor SU(3) symmetry with the standard phase convention. 

Suppressing the transverse-momentum dependence and considering only
the  $S$-wave ($L_{z}=0$) contribution related to the leading-twist
DAs, one finds that Eq.~\eqref{eq:matrixoctet} is equivalent to the
given proton LCWF  
\begin{align}
  |N^{+}(p_{N},1/2)\rangle = \frac{f_{N}}{8\sqrt{6}}
  \int \left[\frac{dx}{\sqrt{x}}\right]_3 
  \bigg{[}  &[V-A]_{N}(x_{i})|u^{\uparrow}u^{\downarrow}
              d^{\uparrow}\rangle
              +[V+A]_{N}(x_{i})|u^{\downarrow}u^{\uparrow}d^{\uparrow}\rangle
              \cr 
&-T_{N}(x_{i})|u^{\uparrow}u^{\uparrow}d^{\downarrow}\rangle\bigg{]},
\end{align}
where the integration measure and the three quark states are
respectively defined as 
\begin{align}
  & \int\left[\frac{dx}{\sqrt{x}}\right]_n:=
    \int  \left[\prod^{n}_{j=1} \frac{dx_{j}}{\sqrt{x_{j}}} \right]
    \delta\left(1-\sum^{n}_{l=1} x_{l}\right), \cr
   & | f^{\sigma_{1}} g^{\sigma_{2}} h^{\sigma_{3}} \rangle =
     \frac{\epsilon^{\alpha_{1}\alpha_{2}\alpha_{3}}}{\sqrt{6}} 
   a^{\dagger}_{\alpha_{1}f\sigma_{1}}(\bm{p}_{1})
   a^{\dagger}_{\alpha_{2}g\sigma_{2}}(\bm{p}_{2})
   a^{\dagger}_{\alpha_{3}h\sigma_{3}}(\bm{p}_{3}) | 0 \rangle.
\label{eq:con1}
\end{align}
The three-quark states carry the color ($\alpha=1,2,3$), flavor
($f=1,2,3=u,d,s$) and spin projection
($\sigma=1,2=\uparrow,\downarrow$) indices. Here the nucleon state,
with the quantum numbers suppressed, is normalized as 
\begin{align}
&\langle N(p) | N(p') \rangle = 2p^{+} (2\pi)^{3} \delta(p^{+}-p'^{+}) \delta^{(2)}(\bm{p}_{\perp}-\bm{p}'_{\perp})
\label{eq:con2}
\end{align}
with the light-cone momenta $p^{\pm}=(p_{0}\pm p_{z})/\sqrt{2}$.

\section{$\Delta$ baryon distribution amplitudes \label{sec:3}}
We now examine the properties of the $\Delta$-baryon DAs. The general decomposition of the vacuum-to-$\Delta$ matrix element of the trilocal quark field operators has been derived in Ref.~\cite{Farrar:1988vz}.
The general decomposition of the corresponding matrix element in the leading twist
is obtained to be 
\begin{align}
  &4\langle 0 | \epsilon^{ijk} u^{i}_{\alpha}(a_1n)
    u^{j}_{\beta}(a_2n)  u^{k}_{\gamma}(a_3n)   |
\Delta^{++} (p_{\Delta},\lambda) \rangle \cr
  &=\lambda^{1/2}_{\Delta} \bigg{[}(\gamma_{\mu}C)_{\alpha \beta}
    \Delta^{\mu}_{\gamma}V_{\Delta}(a_in\cdot p_{\Delta}) + 
    (\gamma_{\mu}\gamma_{5}C)_{\alpha\beta}(
    \gamma_{5}\Delta^{\mu})_{\gamma} A_{\Delta}(a_in\cdot p_{\Delta})
    \cr 
  & \hspace{1.3cm}  - \frac{1}{2}(i\sigma_{\mu\nu}C)_{\alpha\beta}
    (\gamma^{\mu}\Delta^{\nu})_{\gamma}
    T_{\Delta}(a_in\cdot p_{\Delta})\bigg{]}  \cr
  &- f^{3/2}_{\Delta}\bigg{[}(i\sigma_{\mu\nu}C)_{\alpha\beta}
    \left(p^{\mu}\Delta^{\nu}-\frac{1}{2} M_{\Delta} \gamma^{\mu}
    \Delta^{\nu}\right)_{\gamma}\varphi^{3/2}_{\Delta}(a_in\cdot
    p_{\Delta})\bigg{]}, 
\label{eq:matrixdecuplet}
\end{align}
where ${\Delta}^{\mu}(p_{\Delta},\lambda)$ denotes the
Rarita-Schwinger spinor that satisfies  
\begin{align}
  (\slashed{p}_{\Delta}-M_{\Delta})\Delta^{\mu}(p_{\Delta},\lambda)=0,
  \ \ \bar{\Delta}^{\mu}\Delta_{\mu}=-2M_{\Delta}, \ \ \gamma_{\mu} 
  \Delta^{\mu}(p,\lambda)=p^{\mu}_{\Delta} \Delta_{\mu}(p_{\Delta},\lambda)=0.
\end{align}
$f^{1/2}_{\Delta}=\sqrt{2/3}\lambda^{1/2}_{\Delta}/M_{\Delta}$ and $f^{3/2}_{\Delta}$
represent the $\Delta$ baryon decay constants or the DA normalization constants
that is equivalent to the value of  the vacuum-to-$\Delta$ matrix element
at the origin. The Lorentz scalar functions $V_{\Delta},A_{\Delta},T_{\Delta}$ and
$\varphi^{3/2}_{\Delta}$ are therefore normalized to be 
\begin{align}
V_{\Delta}(0)=T_{\Delta}(0)=\varphi^{3/2}_{\Delta}(0)=1, \ \ \ A_{\Delta}(0)=0.
\end{align}
The $\Delta$ baryon DAs satisfy the following symmetries:
\begin{align}
&V_{\Delta}(x_1,x_2,x_3)= V_{\Delta}(x_2,x_1,x_3), \ \ \ A_{\Delta}(x_1,x_2,x_3)=- A_{\Delta}(x_2,x_1,x_3), \cr
&T_{\Delta}(x_1,x_2,x_3)= T_{\Delta}(x_2,x_1,x_3).
\end{align}
and
\begin{align}
T_{\Delta}(x_1,x_2,x_3)= [V-A]_{\Delta}(x_2,x_3,x_1).
\end{align}
Note that $\varphi^{3/2}_{\Delta}(x_1,x_2,x_3)$ is found to be totally symmetric under
the exchange of its variables. Using these symmetries, we can express the $\Delta$ 
DAs in terms of the single DA
$\varphi^{1/2}_{\Delta}(x_1,x_2,x_3)\equiv[V-A]_{\Delta}(x_1,x_2,x_3)$ 
for spin projection $1/2$ and $\varphi^{3/2}_{\Delta}(x_1,x_2,x_3)$
for spin projection $3/2$. 
Suppressing again the representation of the transversal momentum, we arrive at the 
$\Delta^{++}$ LCWF for $\lambda=3/2$
\begin{align}
|\Delta(p_{\Delta},3/2) \rangle&= -\frac{
                                 f^{3/2}_{\Delta}}{24\sqrt{3}}   \int
                                 \left[\frac{dx}{\sqrt{x}}\right]_3
                                 \bigg{[}\varphi^{3/2}_{\Delta}(x_{i})
                                 | u^\uparrow u^\uparrow u^\uparrow
                                 \rangle \bigg{]},  
\label{eq:wf}
\end{align}
and for $\lambda=1/2$
\begin{align}
  |\Delta(p_{\Delta},1/2) \rangle=
  -\frac{f^{1/2}_{\Delta}}{24\sqrt{6}}  \int
  \left[\frac{dx}{\sqrt{x}}\right]_3 
  \bigg{[}&  [V-A]_{\Delta}(x_{i}) | u ^{\uparrow}u^{\downarrow}
            u^{\uparrow} \rangle + [V+A]_{\Delta}(x_{i})|
            u^{\downarrow} u^{\uparrow} u^{\uparrow} \rangle \cr 
&+T_{\Delta}(x_{i}) | u^{\uparrow}u^{\uparrow} u^{\downarrow} \rangle
                                                                    \bigg{]}. 
\end{align}

\section{Light-cone wave function in the chiral quark-soliton
  model\label{sec:4}} 
The effective chiral Lagrangian~\cite{Diakonov:1987ty,
  Christov:1995vm, Diakonov:2002fq} is expressed as
\begin{align}
\mathcal{L}_{\mathrm{eff}} = \overline{\psi}(x)(i\slashed{\partial} -
  M U^{\gamma_{5}})\psi(x), 
\end{align}
where $\psi$ and $U^{\gamma_{5}}$ stand for the quark and chiral
fields respectively. The $M$ 
represents the dynamical quark mass. In principle, the dynamical quark
mass is originally the momentum-dependent one, i.e. $M(p)$ that can be
derived from the zero-mode fermionic solution in the QCD instanton
vacuum~\cite{Diakonov:1985eg, Diakonov:1995qy, Diakonov:2002fq}. This
plays a role of the natural regulator for a quark loop. In this work,
we turn off the momentum dependence of the dynamical  quark mass for
simplicity and introduce an explicit regularization scheme. We use the
Pauli-Villars method. The $SU(2)$ chiral field is defined as: 
\begin{align}
U^{\gamma_{5}} = \frac{1+\gamma_{5}}{2} U +\frac{1-\gamma_{5}}{2}
  U^{\dagger}, \ \ \ U=\exp{(i \pi^{a}(x)\tau^{a})}, 
\end{align}
where $\pi^a(x)$ are the pseudo-Nambu-Goldstone (pNG) boson
fields. The pNG field has a hedgehog symmetry, which is the minimum
generalization of spherical symmetry~\cite{Pauli:1942kwa}:  
\begin{align}
\pi(\bm{x})= n^{a}P(r), \ \ \ \mathrm{with} \ \ \ n^{a}= x^{a}/|\bm{x}|.
\end{align}

The nucleon mass can be derived from the nucleon correlation function,
which is defined as 
\begin{align}
\Pi(T) = \langle 0 | J_{N}(T/2,\bm{0}) J^{\dagger}_{N}(-T/2,\bm{0}) | 0 \rangle,
\end{align}
where the $J_{N}$ is the Ioffe-type current carrying the quantum numbers of the nucleon
\begin{align}
J_{N}(t,\bm{x}) =\frac{1}{N_{c}!} \epsilon^{\beta_{1}...\beta_{N_{c}}} \Gamma^{\{f\}}_{JJ_{3},II_{3}} \Psi_{\beta_{1} f_{1}}(t,\bm{x}) ... \Psi_{\beta_{N_{c}} f_{N_{c}}}(t,\bm{x}).
\end{align}
Here, the $\beta_1,\cdots,\beta_{N_c}$ denote the color indices and $\Gamma^{\{f\}}$
is a matrix with spin-flavor indices $f$.  The $J$ and $I$ represent respectively
the nucleon spin and isospin quantum numbers, and $J_3$ and $I_3$ their third components, respectively. At the large Euclidean time separation, i.e.,
$T \to \infty$, the correlation function yields the nucleon mass $M_{N}$:
\begin{align}
\lim_{T \to \infty} \Pi_{N}(T) \sim e^{-M_{N}T},
\end{align}
where it is evaluated by the functional integral over the quark fields.
Since the functional integral over the chiral field cannot be carried
out exactly, an approximation should be made. In the limit of large
$N_{c}$, we are able to use the saddle point approximation for the
integral over $U$. 
In this approximation, one obtains the classical soliton mass by minimizing
the energy around the saddle point of the pion mean field, which describes
the classical nucleon 
\begin{align}
\frac{\delta M_{N}[U]}{\delta U}\bigg{|}_{U=U_{c}}=0,
\end{align}
and the quantum fluctuation of the pion field is suppressed by $1/N_{c}$. Thus,
we restrict ourselves to the leading order of $N_{c}$.
The self-consistent pionic configuration $U_{c}(\bm{x})$ and the classical soliton mass $M_{N}=1.207 \,\mathrm{GeV}$ were obtained in Ref.~\cite{Diakonov:1988mg}, and the corresponding results are well approximated by the given
arctangent-type profile function  
\begin{align}
P(r) = 2 \arctan\left(\frac{r^{2}_{0}}{r^{2}}\right), \ \ \ r_{0}=
  \frac{0.8}{M}, \ \ \ M=0.345~\mathrm{GeV}, 
\label{eq:profile}
\end{align}
which is depicted in the left panel of Fig.~\ref{fig:1}. Note that the
masses of the nucleon and $\Delta$ baryon are degenerate at the
classical level.

\subsection{Baryon wave functions}
\begin{figure}[htp]
\centering
\includegraphics[scale=0.59]{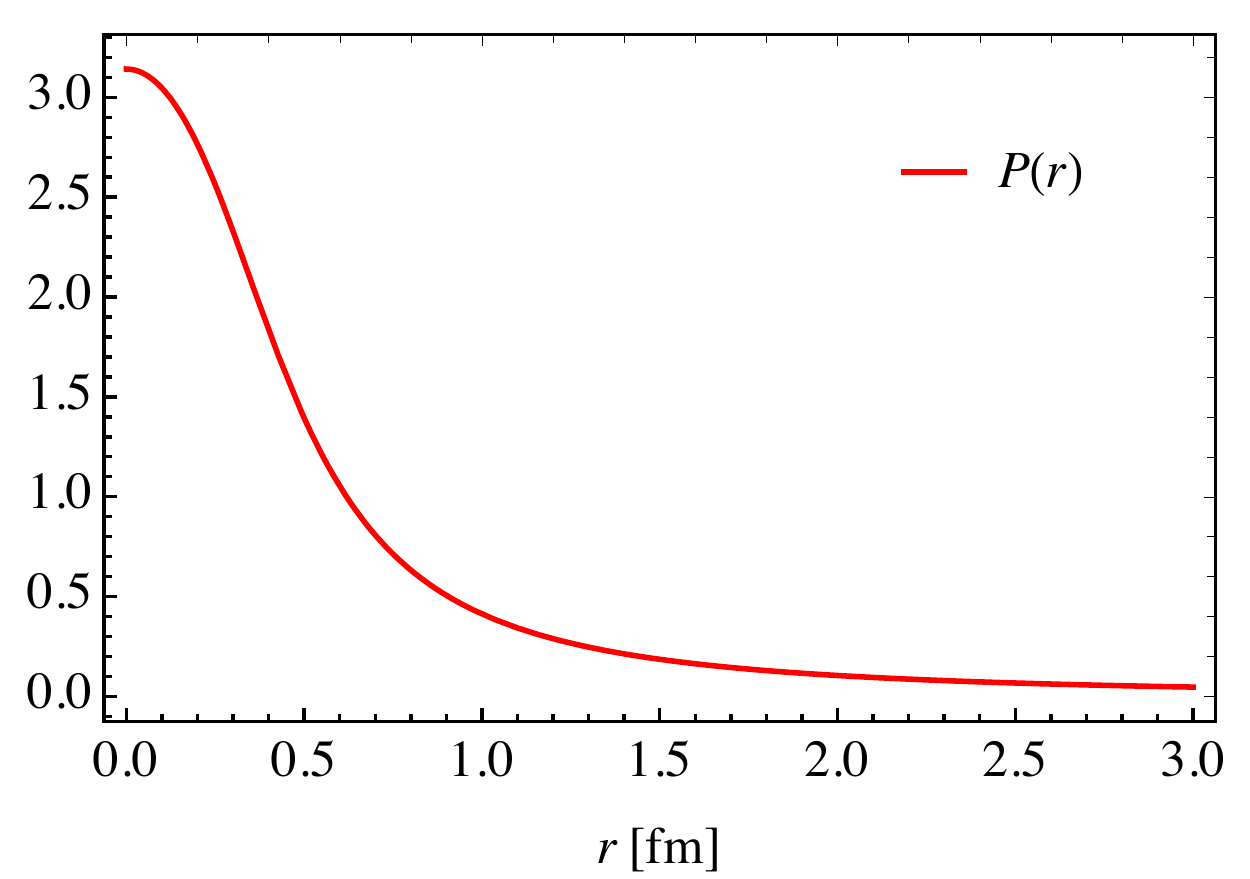}
\includegraphics[scale=0.59]{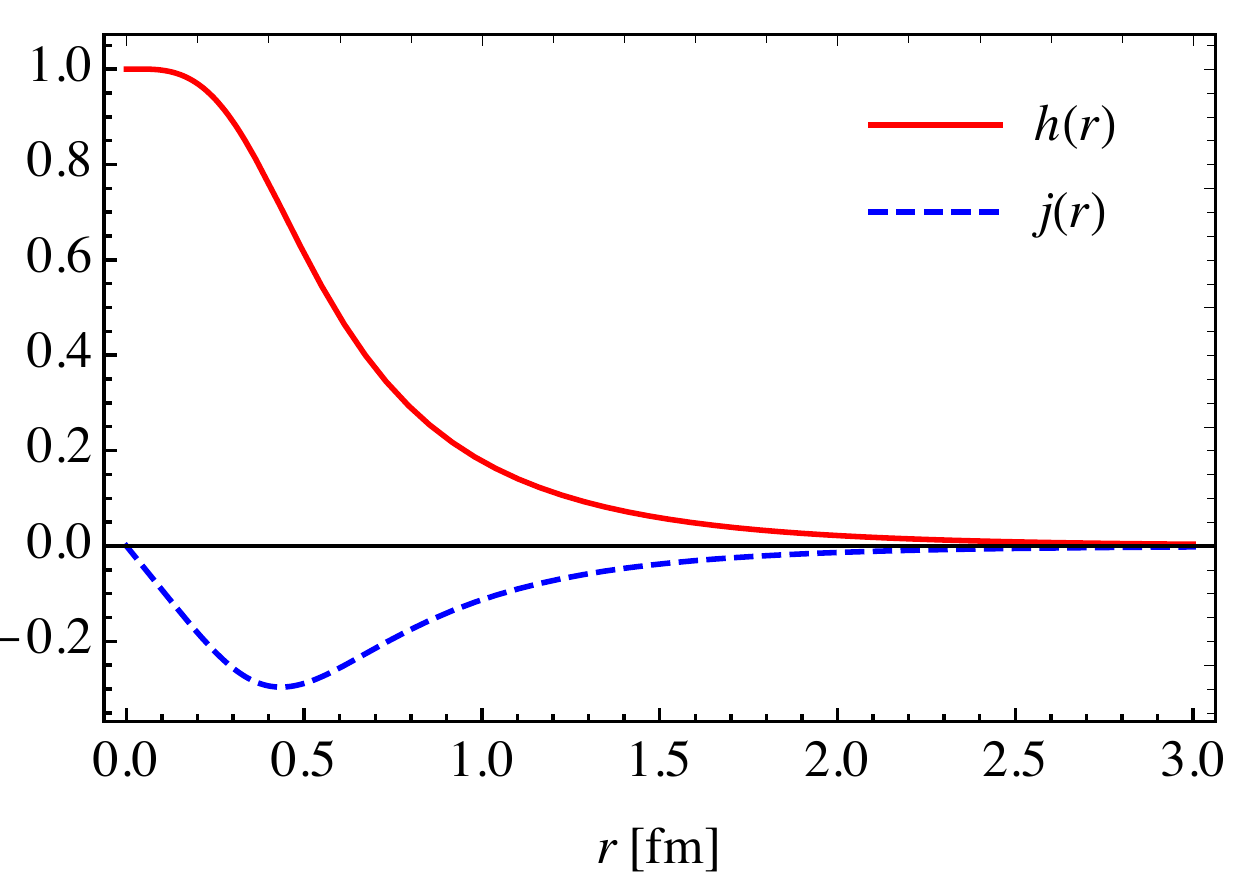}
\caption{The left panel depicts the self-consistent profile function,
  approximated to the arctangent one, as a function of $r$ whereas the
  right panel draws the discrete-level wave functions $h$ and $j$ as a
  function of $r$ with the boundary condition $h(0)=1$ imposed.} 
\label{fig:1}
\end{figure}

To construct the baryon wave function, we need to calculate the
finite-time evolution operator with the definite boundary condition
within the functional integral~\cite{Danilov:1980ez}. In the large
$N_{c}$ limit, the wave function of the classical nucleon (soliton),
completely factorized in the color space, can be obtained by applying
the finite-time evolution operator to any color singlet state of the
$N_{c}$ valence quarks and by taking the finite time $T$ 
to be infinity~\cite{Petrov:2002jr, Diakonov:2004as}:
\begin{align}
| \Psi \rangle &= \prod_{\mathrm{color}} \int
          \frac{d^{3}\bm{p}}{(2\pi)^{3}}F(\bm{p})a^{\dagger}(\bm{p})|\Omega
                 \rangle, 
\label{eq:LCWF1}
\end{align}
where the vacuum wave function or Dirac sea $|\Omega\rangle$ is
given by the quark and anti-quark creation operators 
\begin{align}
  |\Omega \rangle  &= \exp\bigg{[}  \sum_{\mathrm{color}}\int \frac{d^{3}\bm{p}}{(2\pi)^{3}}\frac{d^{3}\bm{p}'}{(2\pi)^{3}} a^{\dagger}(\bm{p})
                     W(\bm{p},\bm{p}') b^{\dagger}(\bm{p}') \bigg{]}|0\rangle,
\label{eq:Coherent_Exp}
\end{align}
with the quark-antiquark pair wave function $W(\bm{p},\bm{p'})$
\begin{align}
W(\bm{p},\bm{p}') &= -i \sqrt{\frac{\varepsilon \varepsilon'}{M^{2}}}   \bar{u}(\bm{p}) G(\bm{p},T,\bm{p'},T) v(\bm{p'}).
\label{eq:pair}
\end{align}
The $\bar{u}$ and $v$ denote the Dirac spinors normalized to be
$\bar{u}u=-\bar{v}v=1$ and $\epsilon=\sqrt{\bm{p}^{2}+M^{2}}$. The creation and  annihilation operators $a^{\dagger}(b^{\dagger})$ 
and $ a(b)$ for a quark (antiquark) satisfy the usual anticommutation relation
\begin{align}
\{ a(\bm{p}'), a^{\dagger}(\bm{p})\}   = (2\pi)^{3} \delta^{(3)}(\bm{p'}-\bm{p}), \ \ \ \{ b(\bm{p}'), b^{\dagger}(\bm{p})\}   = (2\pi)^{3} \delta^{(3)}(\bm{p'}-\bm{p}).
\end{align}
The vacuum state can be annihilated as follows: $a|0\rangle=0$,
$b|0\rangle=0$,  $\langle 0| a^{\dagger}=0$, and $\langle 0|
b^{\dagger}=0$. The quark annihilation-creation operators carry the
color $\alpha$, flavor $f$ and spin projection $\sigma$ indices,
including the three-momentum $\bm{p}$. 
However, we will suppress them except for the momentum from now on. 
The Green function in the presence of the pion mean field in
Eq.~\eqref{eq:pair} and its Fourier transform are defined as 
\begin{align}
&G(\bm{p},T,\bm{p'},T) := \int d^{3}\bm{x}d^{3}\bm{y} e^{-i\bm{p}\cdot
                \bm{x}}e^{-i\bm{p}'\cdot \bm{y}} \,
                G(\bm{x},T,\bm{y},T) \cr 
& \ \ \ \ \ \mbox{ with } \ \ \ \ \   (i\slashed{\partial} - M
                                            U^{\gamma_{5}})G(x,y)=\delta^{(4)}(x-y). 
\label{eq:Green}
\end{align}
$F(\bm{p})$ denotes the valence quark wave function, defined as
\begin{align}
  F(\bm{p}) = \int \frac{d^{3}\bm{p}'}{(2\pi)^{3}}\sqrt{\frac{M}{\varepsilon'}}
  \bigg{[}&   \bar{u}(\bm{p})\gamma^{0}\psi_{\mathrm{lev}}(\bm{p})(2\pi)^{3}
            \delta^{(3)}(\bm{p}-\bm{p}')  \cr
  & -W(\bm{p},\bm{p}')\bar{v}(\bm{p}')
                        \gamma^{0}\psi_{\mathrm{lev}}(-\bm{p}')\bigg{]},
\label{eq:val_WF}
\end{align}
which consists of two parts: the first term corresponds to the
discrete-level wave function whereas the second term arises from the
distortion of this wave function, which is exerted by the vacuum wave
function. In this work, the second term is   ignored bacause of its
complexity. Note that its effects on observables were estimated in
Refs.~\cite{Petrov:2002jr,Lorce:2011dv} in the accuracy of the $3Q$
Fock component and yielded about $10~\%$ contributions. 
The $\psi_{\mathrm{lev}}$ stands for the discrete-level wave function,
satisfying the Dirac equation in the presence of the external chiral
field with the eigenenergies $E_{\mathrm{lev}}$ in the $K^{p} = 0^{+}$
sector with $\bm{K} = \bm{I} + \bm{J}$. $\bm{J}$ and $\bm{I}$ denote
the spin and isospin rotation generators, respectively. It is derived as 
\begin{align}
\psi_{\mathrm{lev}}(r)= \left(\begin{array}{c}  \epsilon^{ji}h(r)  \\
                                -i\epsilon^{jk}(\sigma \cdot
                                \bm{n})^{i}_{k}j(r) \end{array}\right),
  \ \ \ \bigg{\{}\begin{array}{c}  h'+h M \sin P-j(M\cos
                   P+E_{\mathrm{lev}}) = 0  \\ j' +2j/r - j M\sin P -
                   h(M \cos P - E_{\mathrm{lev}})=0 \end{array}, 
\label{eq:levelQ}
\end{align}
where $i=1,2=\uparrow,\downarrow$ denotes the spin projection index
and $j=1,2=u,d$ stands for the isospin index. Inserting the
self-consistent field~\eqref{eq:profile} into Eq.~\eqref{eq:levelQ},
one obtains the eigenenergy $E_{\mathrm{lev}}=0.2\,\mathrm{GeV}$ and
the discrete-level wave function $\psi_{\mathrm{lev}}(r)$ of the
valence quark, which is tightly bound by the self-consistent
fields. In the nonrelativistic limit, the upper component ($L=0$) of
the Dirac bispinor, $h(r)$, is dominant whereas its lower component
($L=1$)  $j(r)\sim0$ is suppressed. $\psi_{\mathrm{lev}}(\bm{p})$ in
Eq.~\eqref{eq:val_WF} is the Fourier transform of
Eq.~\eqref{eq:levelQ}. The discrete-level wave functions are plotted
in the right panel of Fig.~\ref{fig:1}. 

\subsection{Quantization}
The baryon wave function given in Eq.~\eqref{eq:LCWF1} is still a
classical one. While the quantum fluctuations of the pion fields are
completely ignored on account of the $N_c$ counting, the fluctuations
in the direction of the translational and rotational zero modes should
be considered in an exact manner. By this zero-mode quantization, we
are able to construct the baryon state that  acquires the momentum
conservation and the correct quantum numbers~\cite{Diakonov:1987ty,
  Christov:1995vm, Diakonov:2002fq}. As a result,  the quantized
baryon wave function is given by 
\begin{align}
|\Psi^{k}(B) \rangle &=  \int dR B^{*}_{k}(R)
                       \frac{\epsilon^{\alpha_{1}\alpha_{2}\alpha_{3}}}{\sqrt{N_{c}!}}
                       \left[\prod^{N_{c}}_{n=1} \int
                       \frac{d^{3}\bm{p}_{n}}{(2\pi)^{3}}
                       R^{f_{n}}_{j_{n}}
                       F^{j_{n}\sigma_{n}}(\bm{p}_{n})
                       a^{\dagger}_{\alpha_{n}f_{n}\sigma_{n}}(\bm{p}_{n})\right]
                       \cr 
& \times  \exp\left( \int \frac{d^{3}\bm{p}}{(2\pi)^{3}}
                              \frac{d^{3}\bm{p}'}{(2\pi)^{3}}
                              a^{\dagger}_{\alpha f \sigma} (\bm{p})
                              R^{f}_{j}
                              W^{j\sigma}_{j'\sigma'}(\bm{p},\bm{p}')
                              R^{\dagger f'}_{j'} b^{\dagger
                              \alpha f' \sigma'}(\bm{p}')
                              \right) | 0 \rangle,  
\label{eq:LCWF_master}
\end{align}
where $k$ denotes the spin projection of the baryon state. The color
($\alpha=1,2,3$), flavor ($f=1,2,3$), isospin ($j=1,2$) and spin
projection ($\sigma=1,2$) indices suppressed in
Eq.~\eqref{eq:LCWF1} are explicitly restored. 

The higher Fock components are generated by expanding the vacuum wave function $|\Omega\rangle$ to the order of
what we are interested in, i.e., $|\Omega \rangle = e^{\bar{q}q}\sim 1+
\bar{q}q \,+ ...$. Integration over the rotation matrices $R^{f}_{j}$  projects the flavor state of all valence quarks and quark-antiquark out of the vacuum wave
function onto the spin-flavor state $B^{*}(R)$, which describes the baryon octet and
decuplet. While this state can be expressed by the SU(3) Wigner
$D$ functions~\cite{Blotz:1992pw}, which is often evaluated 
in terms of the SU(3) Clebsch–Gordan coefficients, it is more convenient to integrate
over the Haar measure $dR$ to obtain them. 
This method has a great virtue that the symmetries of the quark wave functions are
revealed naturally. The explicit expressions of the baryon rotational wave functions are
given in Appendix~\ref{appendix:a}. Note that the third row of the matrix  $R^{f}_{j}$, $f=3$ makes it possible to pop up the strange quarks both at the valence level and in
the Dirac sea. The physical baryon state is then decomposed into the Fock
components: 
\begin{align}
|\Psi^{k}(B)\rangle &= |\Psi^{(3)k}(B)\rangle + |\Psi^{(5)k}(B)\rangle+ ...,
 \end{align}
where
 \begin{align}
   \langle \Psi^{k}(B) | \Psi^{k}(B) \rangle =\mathcal{N}^{(3)}+
   \mathcal{N}^{(5)} ..., \ \ \ \langle
   \Psi^{(n)k}(B) | \Psi^{(n)k}(B) \rangle 
   = \mathcal{N}^{(n)}. 
\label{eq:normalization}
 \end{align}
 The baryon state is normalized by the factor $\sqrt{\mathcal{N}^{(3)}+
   \mathcal{N}^{(5)} ...}$, which ensures that it should be normalized
 to be unity~\eqref{eq:normalization}. In principle, while all
 possible higher Fock components should be taken into account for
 normalization, it is not possible to consider them in practice. Thus,
 we need to truncate the higher Fock states at a certain level. In
 this work, we include the $3Q$, and $5Q$ components.  

\subsection{Baryon light-cone wave functions}

We are now in a position to derive the LCWFs for the baryon. The LCWF of the baryon
is defined on the light cone. While the stationary saddle-point solution $U_{c}(\bm{x})$ corresponds to the classical baryon at rest, it is necessary to construct the pion mean
field that describes the moving baryon, i.e., $U(t,\bm{x})$, with its velocity $v\to1$.
Since the effective chiral Lagrangian is Lorentz invariant, it is straightforward to boost
the baryon to the light cone. This means that we can easily obtain the discrete-level
wave function and the self-consistent mean field on the light cone by the Lorentz boost. 
Taking the velocity $v\to1$, we derive the valence-quark wave function 
\begin{align}
  &F^{j\sigma}(z,\bm{p}_{\perp}) =  \left(\begin{array}{c c } k_{L}
 f_{\perp}(z,|\bm{k}_{\perp}|) & f_{\parallel}(z,|\bm{k}_{\perp}|) \\ -f_{\parallel}(z,|\bm{k}_{\perp}|) & k_{R}f_{\perp}(z,|\bm{k}_{\perp}|) \end{array}\right)^{j\sigma}
\bigg{|}_{k_{z}=zM_{N}-E_{\mathrm{lev}},\bm{k}_{\perp}=\bm{p}_{\perp}},
\end{align}
with $k_{R,L}=k_{x}\pm i k_{y}$. The two indenpendent functions $f_{\parallel}(z,|\bm{k}_{\perp}|)$ and $f_{\perp}(z,|\bm{k}_{\perp}|)$ are written as
\begin{align}
  & f_{\parallel}(z,|\bm{k}_{\perp}|) 
=\sqrt{\frac{M_{N}}{2\pi}}\left(  h(k) + \frac{k_{z}j(k)}{|\bm{k}|}\right), \ \ \  f_{\perp}(z,|\bm{k}_{\perp}|) =  \sqrt{\frac{M_{N}}{2\pi}} \frac{j(k)}{|\bm{k}|}.
\end{align}
In Fig.~\ref{fig:2}, we illustrate $f_{\parallel}(z,|\bm{k}_{\perp}|)$ and $f_{\perp}(z,|\bm{k}_{\perp}|)$ on the light cone. Note that in the nonrelativistic limit
the function $f^{\mathrm{NR}}_{\parallel}$ is solely responsible for the valence-quark
wave function $F$:
\begin{align}
& f^{\mathrm{NR}}_{\parallel}(z,|\bm{k}_{\perp}|)= \sqrt{\frac{M_{N}}{2\pi}}  h(k), \ \ \  f^{\mathrm{NR}}_{\perp}(z,|\bm{k}_{\perp}|) =  0.
\end{align} 

The creation and annihilation operators for the quarks satisfy the following
anticommutation relation
\begin{align}
\{a^{ \dagger}_{\sigma}(z,\bm{p}_{\perp}), a_{\sigma'}
  (z',\bm{p}'_{\perp})\} =
  \delta_{\sigma\sigma'}\delta(z-z')(2\pi)^{2}\delta^{(2)}
  (\bm{p}_{\perp}-\bm{p}'_{\perp}),  
\end{align}
so that the wave function is properly normalized. It is convenient to rescale
the creation operators to be  $a_{\sigma}(z,\bm{p}_{\perp})
=\sqrt{P_{N}/2\pi}a_{\sigma}(\bm{p})$.
\begin{figure}[htp]
\centering
\includegraphics[scale=0.59]{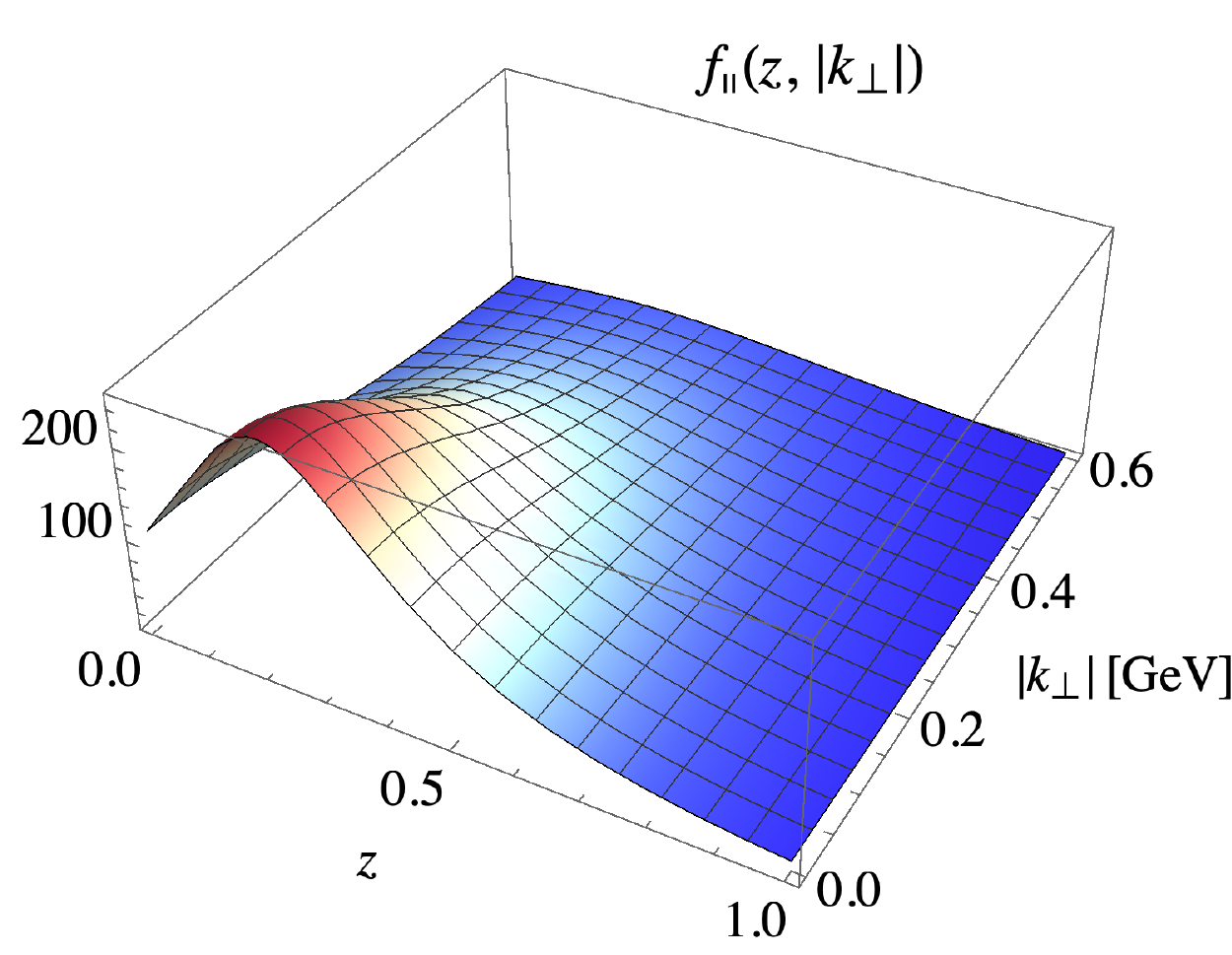}
\includegraphics[scale=0.59]{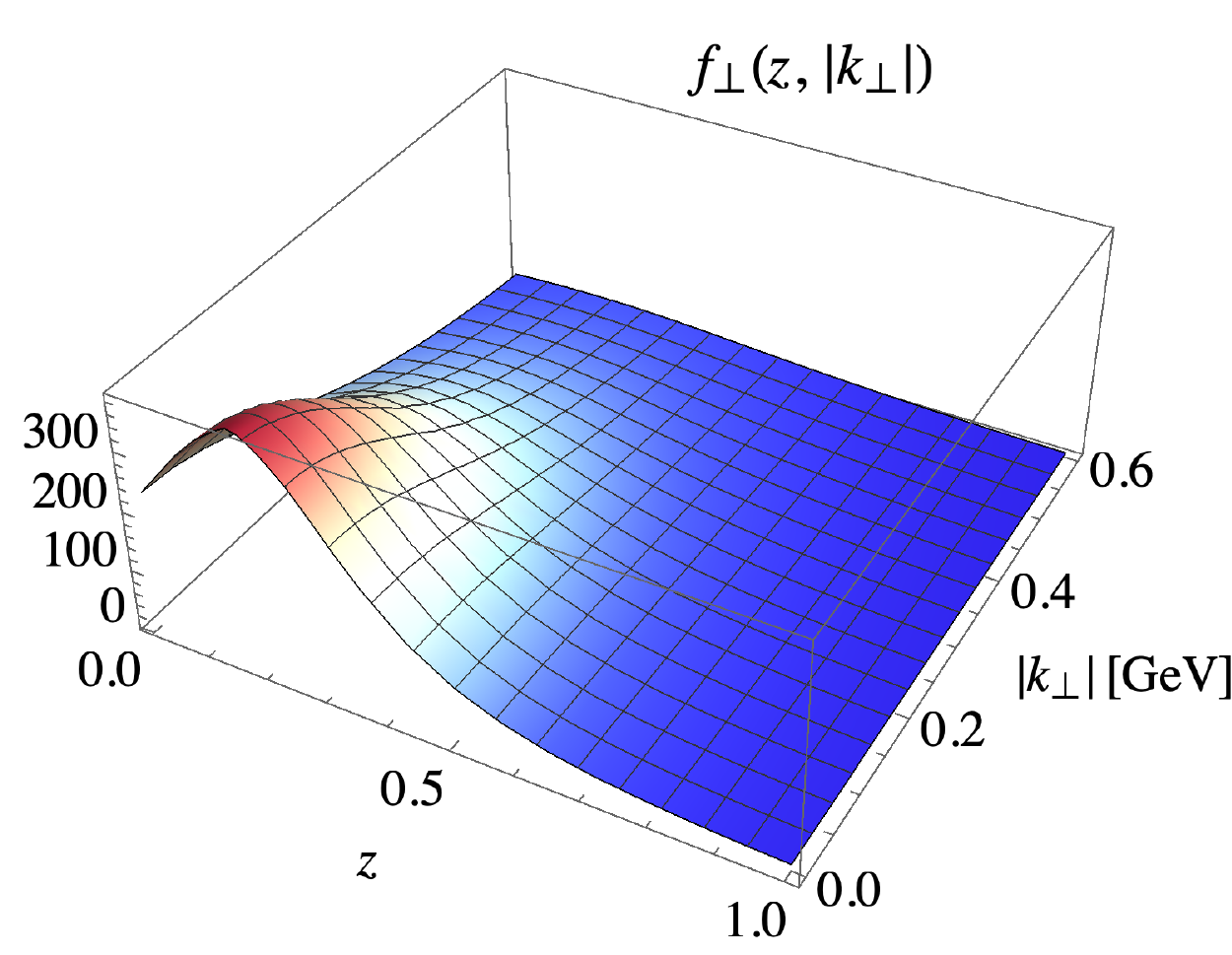}
\caption{The two independent components
  $f_{\parallel}(z,|\bm{k}_{\perp}|)$ and
  $f_{\perp}(z,|\bm{k}_{\perp}|)$ are drawn on the light cone.}  
\label{fig:2}
\end{figure}

The pair wave function $W(\bm{p},\bm{p}')$ is expressed in terms of
the finite-time Green function at an equal time in the presence of the
self-consistent pion field. The Green function is the solution of
Eq.~\eqref{eq:Green}.  The chiral field is decomposed into the scalar
and pseudoscalar fields: 
\begin{align}
U^{\gamma_{5}} = 1+\Sigma + i \Pi \gamma_{5}, 
\end{align}
where $U^{\gamma_5}$ is constrained on the chiral circle 
$(\Sigma+1)^{2}+\bm{\Pi}^{2}=1$. These scalar and pseudoscalar fields in the
momentum space are obtained by the Fourier transforms:
\begin{align}
\Pi(\bm{q})^{j}_{j'} = \int d^{3}\bm{x} e^{-i\bm{q}\cdot \bm{x}} (\bm{n}\cdot\bm{\tau})^{j}_{j'} \sin{P(r)}, \ \ \ \Sigma(\bm{q})^{j}_{j'} = \int d^{3}\bm{x} e^{-i\bm{q}\cdot \bm{x}} \delta^{j}_{j'} (\cos{P(r)}-1)
\end{align}
where the pseudoscalar field $\Pi(\bm{q})$ is a purely imaginary and odd
function, whereas the scalar field $\Sigma(\bm{q})$ is a real and even function.
The explicit forms of these two fields are written as
\begin{align}
&\Pi(\bm{q})^{j}_{j'}=  i \frac{(\bm{q}\cdot
                \bm{\tau})^{j}_{j'}}{|\bm{q}|} \Pi(q), \ \ \
                \Pi(q)=-\int d^{3}\bm{x}~j_{1}(qr) \sin{P(r)}, \cr 
&\Sigma(\bm{q})^{j}_{j'}=  \delta^{j}_{j'} \Sigma(q), \ \ \ \ \  \ \ \
                                                                                                                                                \ \  \  \Sigma(q)=\int d^{3}\bm{x}~j_{0}(qr) (\cos{P(r)}-1).  
\end{align}
$\Sigma{(\bm{q})}$ and $\Pi{(\bm{q})}$ are plotted in Fig~\ref{fig:2}.
\begin{figure}[t]
\centering
\includegraphics[scale=0.59]{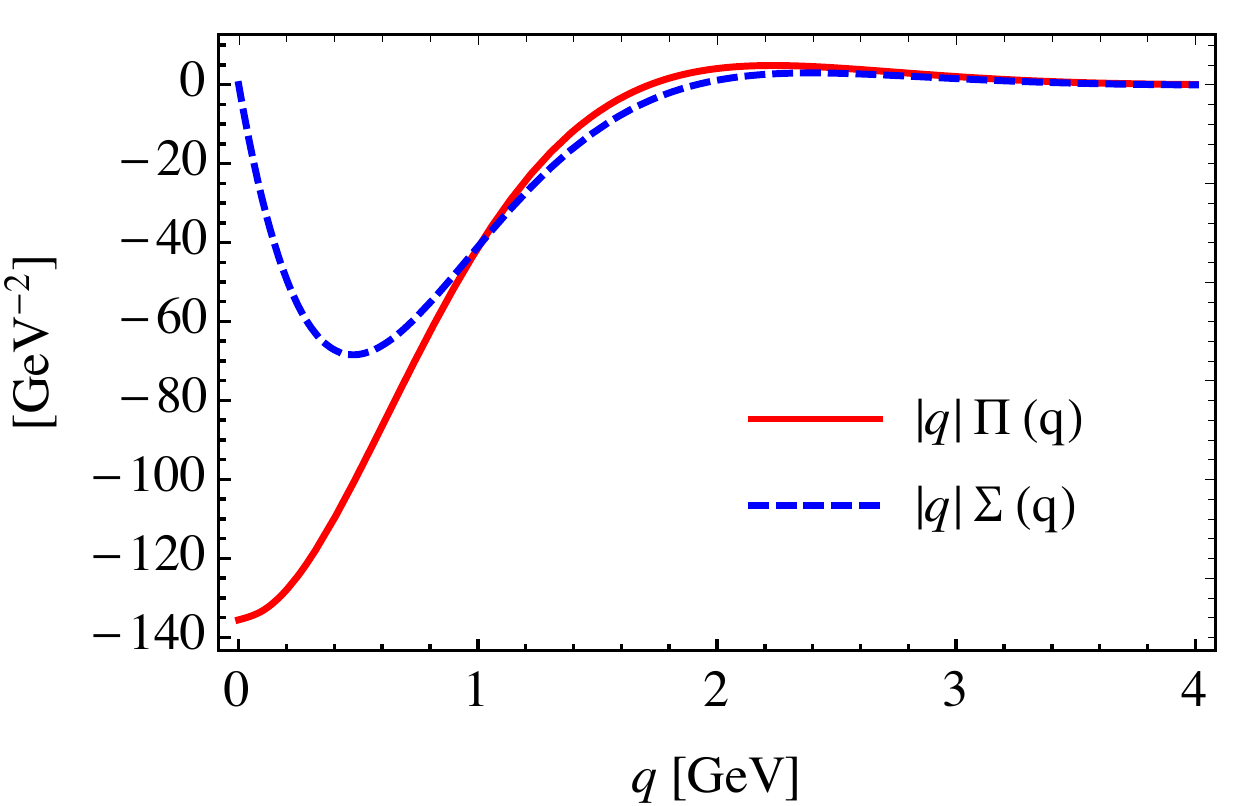}
\caption{Scalar ($|\bm{q}|\Sigma(q)$) and pseudoscalar
  ($|\bm{q}|\Pi(q)$) mean fields are illustrated by the dashed and
  solid curves, respectively.}  
\label{fig:2}
\end{figure}

It is rather complicated to derive the quark-antiquark pair wave function $W(\bm{p},\bm{p'})$ exactly. Hence, we confine ourselves to the
\emph{interpolation approximation} for it. Note that in the three limiting cases of the pion field, i.e., the small pion field $P(r)$,  the slowly varying pion field, and fastly varying pion field, $W(\bm{p},\bm{p'})$ becomes exact. 
Typically, the error between the interpolation approximation
and the exact calculation lies at most within 15~\%. By taking $v\to1$,
the quark-antiquark pair wave function~\eqref{eq:Coherent_Exp} on the light
cone is then obtained to be 
\begin{align}
W^{j \sigma }_{j' \sigma' }(\bm{q},y,\bm{\mathcal{Q}}_{\perp})= &\frac{M_{N}M}{2\pi }\frac{\Sigma^{j}_{j'}(\bm{q})[M(2y-1)\tau_{3} + \bm{\mathcal{Q}}_{\perp} \cdot \bm{\tau}_{\perp}]^{\sigma}_{ \sigma'} + i \Pi^{j}_{j'}(\bm{q})[-M \bm{1} + i \bm{\mathcal{Q}}_{\perp} \times \bm{\tau}_{\perp}]^{\sigma}_{ \sigma'}}{\bm{\mathcal{Q}}^{2}_{\perp}+M^{2}+y(1-y)\bm{q}^{2}},
\label{eq:vac4}
\end{align}
with
\begin{align}
y=\frac{z'}{z+z'}, \ \ \ \bm{\mathcal{Q}}_{\perp}=\frac{z\bm{p}'_{\perp} - z' \bm{p}_{\perp}}{z+z'}, \ \ \ \bm{q}=((\bm{p}+\bm{p}')_{\perp},M_{N}(z+z')).
\label{eq:newvariables}
\end{align}

\subsection{Fock components of the baryon light-cone wave functions}
Various models for the baryon LCWFs mainly focus on the $3Q$ Fock component, 
which is a rather crude approximation in reality. However,
it was found in Refs.~\cite{Diakonov:2005ib, Lorce:2007as} that the higher Fock
components are nonnegligible and come into significant play in studying the various  features of the nucleon. The $5Q$ Fock component provides at least 20~\% contribution to
observables. Thus, we will derive both the $3Q$ and $5Q$ wave
functions for the baryon octet and decuplet.
\subsubsection{$3Q$ Fock component}
Expanding Eq.~\eqref{eq:LCWF_master} and taking its leading term, 
we derive the baryon LCWF of the $3Q$ component
\begin{align}
|\Psi^{(3)k}(B) \rangle &= T(B)^{f_{1}f_{2}f_{3}}_{j_{1}j_{2}j_{3},k} \frac{\epsilon^{\alpha_{1}\alpha_{2}\alpha_{3}}}{\sqrt{N_{c}!}}   \prod^{N_{c}}_{n=1} \int \frac{d^{3}\bm{p}_{n}}{(2\pi)^{3}}  F^{j_{n}\sigma_{n}}(\bm{p}_{n})a^{\dagger}_{\alpha_{n}f_{n}\sigma_{n}}(\bm{p}_{n}) | 0 \rangle.
\label{eq:3QWF}
\end{align}
Each of three valence quarks is rotated by the matrix $R^{f}_{j}$ and projected
onto the spin-flavor baryon state $B^*(R)$ by integrating over $R$. The
short-handed notation for this group integral is defined as 
\begin{align}
  T(B)^{f_{1}f_{2}f_{3}}_{j_{1}j_{2}j_{3},k} :=
  \int dR B^{*}_{k}(R)R^{f_{1}}_{j_{1}} R^{f_{2}}_{j_{2}} R^{f_{3}}_{j_{3}}.
\end{align}
We refer to Refs.~\cite{Diakonov:2005ib, Lorce:2006nq,Lorce:2007as,
  Lorce:2007xax} for details. 

The normalization $\mathcal{N}^{(3)}$ of the baryon LCWF from
Eq.~\eqref{eq:normalization} is obtained by contracting the creation
and annihilation operators, which is written as
\begin{align}
\mathcal{N}^{(3)}(B) &= 6 T(B)^{f_1f_2f_3}_{j_1j_2j_3,k}
                       T(B)_{f_1f_2f_3}^{j'_1j'_2j'_3,k} \cr 
  &\times \int [d\bm{p}]_n
    F^{j_1\sigma_1}(\bm{p}_1)F^{j_2\sigma_2}(\bm{p}_{2})F^{j_3\sigma_3}
    (\bm{p}_{3}) F^{\dagger}_{j'_1\sigma_1}(\bm{p}_1)
    F^{\dagger}_{j'_2\sigma_2}(\bm{p}_2)
    F^{\dagger}_{j'_3\sigma_3}(\bm{p}_3), 
\label{eq:3Qnor}
\end{align}
where the integration measure for $nQ$ state is given by
\begin{align}
  \int [d\bm{p}]_n = \int dz_1\cdots dz_n\, \delta\left(\sum^{n}_{l=1}
  z_{l}-1   \right)\,  \int \left( \prod_{i=1}^n
  \frac{d^{2}\bm{p}_{i\perp}}{(2\pi)^{2}} \right)
  (2\pi)^{2}\delta^{(2)}\left(\sum^{n}_{l=1}
  \bm{p}_{l\perp}\right).
\label{eq:Imeasure}
\end{align}
Here $F(\bm{p}_{i}):= F(z_{i},\bm{p}_{i\perp})$ denote the quark wave functions. 
To compute $\mathcal{N}^{(3)}(B)$, it is convenient to introduce the
probability distribution  $\Phi(z,\bm{q}_{\perp})$. It also helps to
compute the normalization for the $5Q$ component and physical
observables. Moreover, $\Phi(z,\bm{q}_{\perp})$ provides information
on how the valence quarks leave the longitudinal momentum fraction of
the baryon and transverse momentum to the quark-antiquark pair wave
function. Thus, it is given as a function of the longitudinal momentum
fraction $z=q_{z}/M_{N}$ and transverse momentum $\bm{q}_\perp$: 
\begin{align}
  \Phi(z,\bm{q}_{\perp}) &= \int dz_1 dz_2 dz_3\, \delta\left(\sum^{3}_{l=1}
                           z_{l}+z-1\right) \int \left(\prod_{i=1}^3
                           \frac{d^{2}\bm{p}_{i\perp}}{(2\pi)^{2}}\right) \cr
&\times  (2\pi)^{2}\delta^{(2)}\left(\sum^{3}_{l=1}\bm{p}_{l\perp} +\bm{q}_{\perp}\right) D(\bm{p}_1,\bm{p}_2,\bm{p}_3),
\label{eq:valence_prob}
\end{align}
where $D(\bm{p}_1,\bm{p}_2,\bm{p}_3)$ is defined as 
\begin{align}
D(\bm{p}_1,\bm{p}_2,\bm{p}_3) :=\bigg{[}f^{2}_{\parallel}(\bm{p}_1)+p_{1R}p_{1L}f^{2}_{\perp}(\bm{p}_1)\bigg{]}\bigg{[}f^{2}_{\parallel}(\bm{p}_2)+p_{2R}p_{2L}f^{2}_{\perp}(\bm{p}_2)\bigg{]}\bigg{[}f^{2}_{\parallel}(\bm{p}_3)+p_{3R}p_{3L}f^{2}_{\perp}(\bm{p}_3)\bigg{]}
\label{eq:valence_prob_func}
\end{align}
with the discrete-level wave functions $f_{\perp,\parallel}(\bm{p}_i) :=
f_{\perp,\parallel}(z_{i},|\bm{p}_{i\perp}|)$.
In the non-relativistic limit, $D(\bm{p}_1,\bm{p}_2,\bm{p}_3)$ is reduced to
\begin{align}
  D^{\mathrm{NR}}(\bm{p}_1,\bm{p}_2,\bm{p}_3)=
  \left[f^{\mathrm{NR}}_{\parallel}(\bm{p}_1)f^{\mathrm{NR}}_{\parallel}(\bm{p}_2)
  f^{\mathrm{NR}}_{\parallel}(\bm{p}_3)\right]^{2}.  
\end{align}

\begin{figure}[htp]
\centering
\includegraphics[scale=0.59]{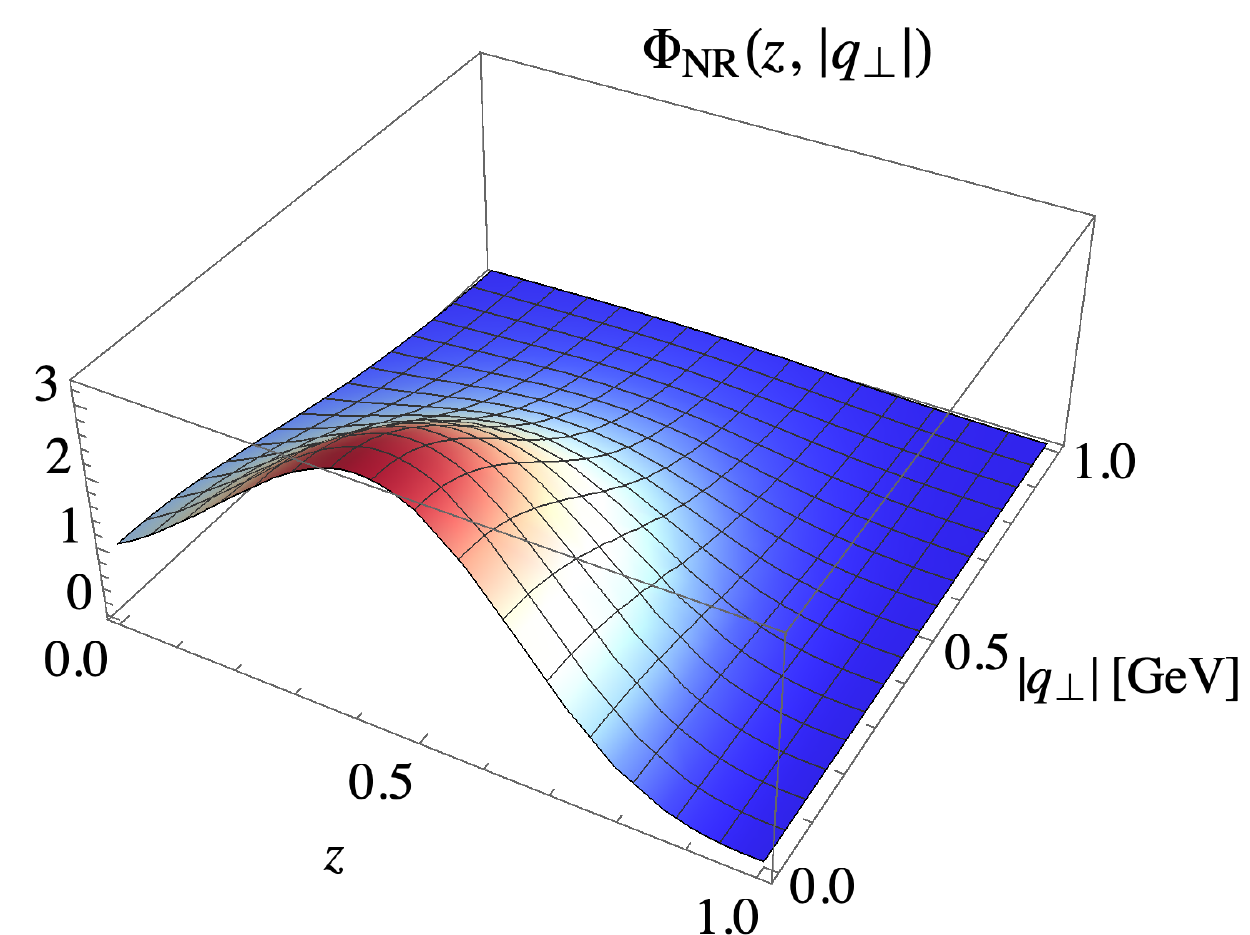}
\includegraphics[scale=0.59]{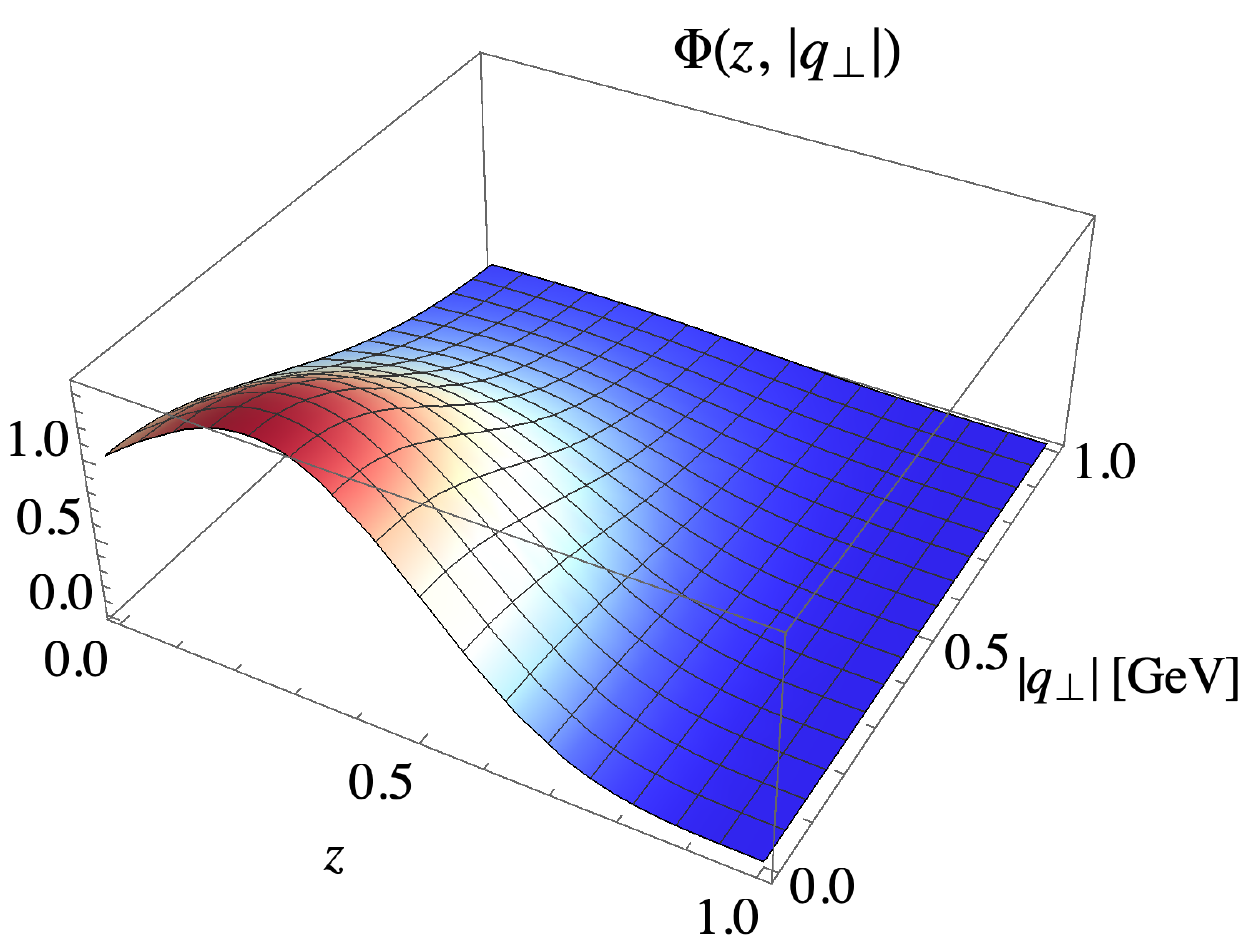}
\caption{The left (right) panel depicts the probability distribution $\Phi(x,\bm{q}_{\perp})$ that the valence quarks leave the longitudinal momentum fraction $z$ of the baryon momentum and transverse momentum $\bm{q}_{\perp}$ to the quark-antiquark pair wave functions in the case of the non-relativistic (relativistic) picture.}
\label{fig:4}
\end{figure}

Since in the $3Q$ wave function there is no additional quark-antiquark pair, the normalization is thus proportional to $\Phi(0,0)$:
\begin{align}
\mathcal{N}^{(3)}(B_{8})=\frac{3}{2}\Phi(0,0), \ \ \ \mathcal{N}^{(3)}_{1/2}(B_{10})=\mathcal{N}^{(3)}_{3/2}(B_{10})=\frac{3}{5}\Phi(0,0).
\end{align}
Note that the normalization of the discrete-level wave functions $f_{\perp}$ and $f_{\parallel}$ are arbitrary. We choose it to be $\Phi(0,0)=1$.
The corresponding results for the probability distribution are plotted in
Fig.~\ref{fig:4} for both the non-relativistic and relativistic cases. 

\subsubsection{$5Q$ Fock component}
We now compute the $5Q$ Fock component of the baryon LCWF.  Note that
it has an important contribution to various baryonic observables,  
e.g., the axial charge $g_{A}$, the DA normalization constant $f_{N}$,
and so on. Expanding  Eq.~\eqref{eq:LCWF_master} to the linear order
in the  $Q\bar{Q}$ pair, we obtain the $5Q$ Fock component of the
baryon LCWF:  
\begin{align}
|\Psi^{(5)k}(B) \rangle &=
                          T(B)^{f_{1}f_{2}f_{3}f_4,j_5}_{j_{1}j_{2}j_{3}j_4,f_5,k}
                          \frac{\epsilon^{\alpha_{1}\alpha_{2}\alpha_{3}}}{\sqrt{N_{c}!}}
                          \left[ \prod^{N_{c}}_{n=1} \int
                          \frac{d^{3}\bm{p}_{n}}{(2\pi)^{3}}
                          F^{j_{n}\sigma_{n}}(\bm{p}_{n})
                          a^{\dagger}_{\alpha_{n}f_{n}\sigma_{n}}(\bm{p}_{n}) 
                          \right]  \cr 
& \times  \int \left(\prod_{i=4}^5
                               \frac{d^{3}\bm{p}_{i}}{(2\pi)^{3}}\right)
   a^{\dagger}_{\alpha_{4}   f_{4} \sigma_{4}} (\bm{p_{4}})
     \delta^{\alpha_4}_{^{\alpha_5}}
                                          W^{j_{4}\sigma_{4}}_{j_{5}\sigma_{5}}(
                                          \bm{p_{4}},\bm{p_{5}}) 
    b^{\dagger \alpha_{5} f_{5}  \sigma_{5}}(\bm{p_{5}})  | 0
                                          \rangle,  
\end{align}
where the indices with $1,2,3$ represent the valence quarks, whereas those with
$4,5$ stand for the quark and antiquark of $Q\bar{Q}$ pair,
respectively. The color indices (1,2,3) for the valence quarks are
antisymmetric and those for the quark-antiquark pair (4,5) become a
color singlet. The group integral over the valence quarks and
$Q\bar{Q}$-pair rotation matrices with the spin-flavor state
$B^{*}_{k}$ is expressed as 
\begin{align}
  T(B)^{f_{1}f_{2}f_{3}f_4,j_5}_{j_{1}j_{2}j_{3}j_4,f_5,k}:= \int dR B^{*}_{k}(R)R^{f_{1}}_{j_{1}}
  R^{f_{2}}_{j_{2}} R^{f_{3}}_{j_{3}} R^{f_{4}}_{j_{4}} R^{\dagger j_{5}}_{f_{5}}.
\end{align}

The normalization constant $\mathcal{N}^{(5)}$ is obtained by contracting the
creation and annihilation operators for the valence quarks and $Q\bar{Q}$ pair.
In the presence of the quark-antiquark creation operators, two
additional contractions (exchange and direct contributions) are allowed.
So, we get the following general expression for the normalization of the $5Q$
component:
\begin{align}
\mathcal{N}^{(5)}(B) &= 18 T(B)^{f_{1}f_{2}f_{3}f_4,j_5}_{j_{1}j_{2}j_{3}j_4,f_5,k} T(B)^{j'_{1}j'_{2}j'_{3}j'_4,f_5,k}_{f_{1}f_{2}g_{3}g_4,j'_5} \int [d\bm{p}]_{5} \cr
&\times  F^{j_1\sigma_1}(\bm{p}_1)F^{j_2\sigma_2}(\bm{p}_2)F^{j_3\sigma_3}(\bm{p}_3) W^{j_4 \sigma_4}_{j_5\sigma_5} F^{\dagger}_{j'_1\sigma_1}(\bm{p}_1) F^{\dagger}_{j'_2\sigma_2}(\bm{p}_2) \cr
&\times \bigg{[}F^{\dagger}_{j'_3\sigma_3}(\bm{p}_3)W^{\dagger j'_5 \sigma_{5}}_{j'_4\sigma_4}(\bm{p}_{4},\bm{p}_{5}) \delta^{g_{3}}_{f_{3}}\delta^{g_{4}}_{f_{4}} - F^{\dagger}_{j'_3\sigma_4}(\bm{p}_4)W^{\dagger j'_5 \sigma_{5}}_{j'_4\sigma_3}(\bm{p}_{3},\bm{p}_{5}) \delta^{g_{3}}_{f_{4}}\delta^{g_{4}}_{f_{3}}\bigg{]}
\label{eq:5Qnor}
\end{align}
with the integration measure defined in Eq.~\eqref{eq:Imeasure}.
The first and second terms
in the bracket are called the direct and exchange contributions, respectively.
Since it was found that the exchange contributions to the normalization
$\mathcal{N}^{(5)}$ are negligibly small as shown in
Refs.~\cite{Lorce:2006nq,Lorce:2007xax}, we ignore them.  
Note that in the case of the direct contributions the $Q\bar{Q}$-pair momenta are
almost decoupled from the valence quark momenta. Thus, we can define the
following function with the valence quarks and $Q\bar{Q}$ pair factorized:  
\begin{align}
K_{J} := \frac{M^{2}}{2\pi} \int \frac{d^{3}\bm{q}}{(2\pi)^{3}} \Phi\left(\frac{q_{z}}{M_{N}},\bm{q}_{\perp}\right) \theta(q_{z}) q_{z} G_{J}(q_{z},\bm{q}_{\perp}),
\label{eq:5Q1}
\end{align}
where $G_{J}$ are quark-antiquark probability distributions with $J=\pi\pi$, $\sigma\sigma$, and $33$, which arises from the quark-antiquark loops:
\begin{align}
&G_{\sigma\sigma}(q_{z},\bm{q}_{\perp})= \Sigma^{2}(\bm{q}) \int^{1}_{0} dy \int \frac{d^{2}\bm{\mathcal{Q}}_{\perp}}{(2\pi)^{2}}\left[\frac{M^{2}(2y-1)^{2} + \bm{\mathcal{Q}}^{2}_{\perp} }{\left(\bm{\mathcal{Q}}^{2}_{\perp} + M^{2} + y(1-y)\bm{q}^{2}\right)^{2}} - (M-M_{\mathrm{PV}}) \right], \cr
&G_{\pi\pi}(q_{z},\bm{q}_{\perp})= \Pi^{2}(\bm{q}) \int^{1}_{0} dy \int \frac{d^{2}\bm{\mathcal{Q}}_{\perp}}{(2\pi)^{2}}\left[\frac{M^{2}+\bm{\mathcal{Q}}^{2}_{\perp} }{\left(\bm{\mathcal{Q}}^{2}_{\perp} + M^{2} + y(1-y)\bm{q}^{2}\right)^{2}} - (M-M_{\mathrm{PV}}) \right], \cr
&G_{33}(q_{z},\bm{q}_{\perp})= \frac{q^{2}_{z}}{\bm{q}^{2}}\Pi^{2}(\bm{q}) \int^{1}_{0} dy \int \frac{d^{2}\bm{\mathcal{Q}}_{\perp}}{(2\pi)^{2}}\left[\frac{M^{2}+\bm{\mathcal{Q}}^{2}_{\perp} }{\left(\bm{\mathcal{Q}}^{2}_{\perp} + M^{2} + y(1-y)\bm{q}^{2}\right)^{2}} - (M-M_{\mathrm{PV}}) \right]. \cr
\label{eq:5Q2}
\end{align}
Since $G_J$ are divergent logarithmically, we introduce the Pauli-Villars regulator
to tame them. After some straightforward manipulations, one arrives at the baryon 
normalization constant $\mathcal{N}^{(5)}$ expressed in terms of $K_J$ 
\begin{align}
\mathcal{N}^{(5)}(B_{8}) &= \frac{3}{10}(11K_{\pi \pi} + 23K_{\sigma \sigma}), \cr
  \mathcal{N}^{(5)}_{1/2}(B_{10}) &= \frac{3}{20}(11K_{\pi \pi} + 6 K_{33} +
    17K_{\sigma \sigma}), \cr
    \mathcal{N}^{(5)}_{3/2}(B_{10})
    &= \frac{3}{20}(15K_{\pi \pi} - 6 K_{33} + 17K_{\sigma \sigma}).
\end{align}

\subsection{Distribution amplitudes}
So far, we have defined the baryon LCWF normalized to be unity in the
instant form for simplicity. However, it is more convenient to introduce the LCWF given as a function of the light-cone variables, since it can be directly applied to
many baryonic observables such as the electromagnetic form factors of the
nucleon~\cite{Lepage:1980fj, Chernyak:1983ej}. Thus, we reexpress it in the
covariantly normalized form in terms of the light-cone variables, keeping the
physics intact:
\begin{align}
  |N(p_{N},\lambda)\rangle &= c^{\mathcal{R},J_z}_{0}  T(B)^{fgh}_{j_{1}j_{2}j_{3},k} \int \left[\frac{dx}{\sqrt{x}}\right]_3 \cr
  & \times \int [dk_\perp]_{3}
    F^{j_1\sigma_1}(\bm{k}_1)F^{j_2\sigma_2}(\bm{k}_2)F^{j_3\sigma_3}(\bm{k}_3)  | f^{\sigma_{1}} g^{\sigma_{2}} h^{\sigma_{3}} \rangle,
\label{eq:cov_WF}
\end{align}
where the fermion states are normalized to be Eqs.~\eqref{eq:con1}
and~\eqref{eq:con2} and the normalization of the LCWF is expressed as
\begin{align}
c^{\mathcal{R},J_z}_{0}=\sqrt{  \frac{(4\pi)^{2}}{\mathcal{N}^{(3)}+\mathcal{N}^{(5)}...}}.
\end{align}
The well-known abbreviated notations of the integration measures are adopted as
follows:  
\begin{align}
&\int\left[\frac{dx}{\sqrt{x}}\right]_n:=\int  \left[\prod^{n}_{j=1} \frac{dx_{j}}{\sqrt{x_{j}}} \right] \delta\left(1-\sum^{n}_{l=1} x_{l}\right), \cr & \int[dk_\perp]_{n}:= \int \left[\prod^{n}_{j=1} \frac{d\bm{k}_{j\perp}}{2(2\pi)^{3}}\right] 2(2\pi)^{3} \delta^{(2)}\left(\sum^{n}_{l=1} \bm{k}_{l\perp}\right).
\end{align}
Applying the three quark field operators to the baryon LCWF~\eqref{eq:cov_WF}
in Eq.~\eqref{eq:matrixoctet}, one can extract respectively the DAs
for the nucleon and $\Delta$ baryon
\begin{align}
&f_{N}\varphi_{N} = 24\sqrt{6} c^{\bm{8},1/2}_{0} \int [dk_{\perp}] T^{112}_{j_{1}j_{2}j_{3},1}  F^{j_{1}1}(\bm{k}_{1})F^{j_{2}2}(\bm{k}_{2})F^{j_{3}1}(\bm{k}_{3}), \cr
&f^{1/2}_{\Delta}\varphi^{1/2}_{\Delta} = -24\sqrt{6} c^{\bm{10},1/2}_{0} \int [dk_{\perp}] T^{111}_{j_{1}j_{2}j_{3},1}  F^{j_{1}1}(\bm{k}_{1})F^{j_{2}2}(\bm{k}_{2})F^{j_{3}1}(\bm{k}_{3}),\cr
&f^{3/2}_{\Delta}\varphi^{3/2}_{\Delta} = -24\sqrt{3} c^{\bm{10},3/2}_{0} \int [dk_{\perp}] T^{111}_{j_{1}j_{2}j_{3},1}  F^{j_{1}1}(\bm{k}_{1})F^{j_{2}1}(\bm{k}_{2})F^{j_{3}1}(\bm{k}_{3}). 
\end{align}
Summing over the isospin indices, we obtain the explicit expressions for the
DAs
\begin{align}
&f_{N}\varphi_{N} = 4\sqrt{3} c^{\bm{8},1/2}_{0} \int [dk_{\perp}] \bigg{[}f_{\parallel}(\bm{\bm{k}_{1}})f_{\parallel}(\bm{\bm{k}_{2}})f_{\parallel}(\bm{\bm{k}_{3}}) \cr
& \hspace{3cm}-k_{1L}k_{2R}f_{\perp}(\bm{\bm{k}_{1}})f_{\perp}(\bm{\bm{k}_{2}})f_{\parallel}(\bm{\bm{k}_{3}})+2k_{3L}k_{2R}f_{\perp}(\bm{\bm{k}_{3}})f_{\perp}(\bm{\bm{k}_{2}})f_{\parallel}(\bm{\bm{k}_{1}})\bigg{]}, \cr
&f^{1/2}_{\Delta}\varphi^{1/2}_{\Delta} = -\frac{24}{\sqrt{5}} c^{\bm{10},1/2}_{0} \int [dk_{\perp}] \bigg{[}f_{\parallel}(\bm{\bm{k}_{1}})f_{\parallel}(\bm{\bm{k}_{2}})f_{\parallel}(\bm{\bm{k}_{3}}) \cr
& \hspace{3cm}-k_{1L}k_{2R}f_{\perp}(\bm{\bm{k}_{1}})f_{\perp}(\bm{\bm{k}_{2}})f_{\parallel}(\bm{\bm{k}_{3}})-k_{3L}k_{2R}f_{\perp}(\bm{\bm{k}_{3}})f_{\perp}(\bm{\bm{k}_{2}})f_{\parallel}(\bm{\bm{k}_{1}})\bigg{]}, \cr
&f^{3/2}_{\Delta}\varphi^{3/2}_{\Delta} = 12 \sqrt{\frac{6}{5}} c^{\bm{10},3/2}_{0} \int [dk_{\perp}] \bigg{[}f_{\parallel}(\bm{\bm{k}_{1}})f_{\parallel}(\bm{\bm{k}_{2}})f_{\parallel}(\bm{\bm{k}_{3}}) \bigg{]}. 
\label{eq:DAs_Model}
\end{align}
If the LCWF is normalized to the $3Q$ components in the non-relativistic limit, then 
the normalization constants $f^{(3)}_{N},f^{1/2,(3)}_{\Delta},f^{3/2,(3)}_{\Delta}$ become
\begin{align}
\bigg{|}\frac{f^{1/2,(3)}_{\Delta}}{f^{(3)}_{N}}\bigg{|}= \sqrt{6}, \ \ \ \bigg{|}\frac{f^{3/2,(3)}_{\Delta}}{f^{(3)}_{N}}\bigg{|}= 3, \ \ \ \bigg{|}\frac{f^{1/2,(3)}_{\Delta}}{f^{3/2,(3)}_{\Delta}}\bigg{|}= \sqrt{\frac{2}{3}}.
\label{eq:relation_NR}
\end{align}
The expressions for the distribution amplitudes $A,V$ and $T$ are listed
in Appendix~\ref{appendix:b}.

\section{Numerical results and discussion \label{sec:5}}
To compute the baryon LCWFs, we have to fix the parameters of the $\chi$QSM. 
The Pauli-Villars cutoff mass is fixed to be $M_{\mathrm{PV}}=557\,\mathrm{MeV}$
by reproducing the pion decay constant $F_{\pi}=93\,\mathrm{MeV}$. 
As mentioned previously, the dynamical quark mass $M$ is originally momentum-dependent, which comes from the instanton zero mode. Using the
standard values for the instanton ensemble, i.e. the instanton density
$N/V=(1\,\mathrm{fm})^{-1}$ and the average size of the instanton
$\rho=\frac13$ fm, one obtains $M=M(0)=345 \, \mathrm{MeV}$
at the zero quark virtuality. We will use this value in this work.  
By putting all the obtained functions into Eq.~\eqref{eq:5Q1}, the
numerical results for $K_{\pi \pi}$, $K_{\sigma \sigma}$ and $K_{33}$ are
evaluated to be 
\begin{align}
K^{\mathrm{NR}}_{\sigma \sigma}&= 0.059, \ \ \ K^{\mathrm{NR}}_{\pi \pi}= 0.128, \ \ \ K^{\mathrm{NR}}_{33}= 0.077, \cr
K_{\sigma \sigma}&= 0.029, \ \ \ K_{\pi \pi}= 0.075, \ \ \ K_{33}= 0.041.
\end{align}
Employing these results, we determine the non-relativistic and relativistic normalizations for both the baryon octet and decuplet
\begin{align}
& \mathcal{N}^{(5)}(B_{8}) = 0.45, \ \ \ \mathcal{N}^{(5)}_{1/2}(B_{10}) = 0.23, \ \ \ \mathcal{N}^{(5)}_{3/2}(B_{10}) = 0.21.
\label{eq:normres}
\end{align}
We list in Table~\ref{tab:1} each contribution of the $3Q$ and $5Q$ components respectively to the normalization constant for the baryon octet and decuplet in the nonrelativistic case. In Table~\ref{tab:2}, we list them in the relativistic case. 
\begin{table}[htp]
\centering
\setlength{\tabcolsep}{5pt}
\renewcommand{\arraystretch}{1.5}
\begin{tabular}{ c | c c } 
\hline
\hline
$B_{\mathcal{R},J_{z}}$  & $3Q \equiv
                           \frac{\mathcal{N}^{(3)}_{J_{z}}(B_{\mathcal{R}})}{
                           \mathcal{N}^{(3)}_{J_{z}}(B_{\mathcal{R}})
                           +\mathcal{N}^{(5)}_{J_{z}}(B_{\mathcal{R}})}$  
  & $5Q \equiv
    \frac{\mathcal{N}^{(5)}_{J_{z}}(B_{\mathcal{R}})}{
    \mathcal{N}^{(3)}_{J_{z}}(B_{\mathcal{R}})+\mathcal{N}^{(5)}_{J_{z}}(
    B_{\mathcal{R}})}$ 
  \\ 
\hline
$B_{8,1/2}$ & $64\%$ & $36\%$ \\
$B_{10,1/2}$ & $58\%$ & $42\%$ \\
$B_{10,3/2}$ &$62\%$ & $38\%$ \\
\hline
\hline
\end{tabular}
\caption{$3Q$ and $5Q$ fractions of the normalization constants for
  the baryon octet and decuplet in the nonrelativistic
  case. $\mathcal{R}$ stands for the SU(3) representation of the
  baryon multiplet. The spin projection of the corresponding baryon  
  multiplet is denoted by $J_{z}$.} 
\label{tab:1}
\end{table}
\begin{table}[htp]
\centering
\setlength{\tabcolsep}{5pt}
\renewcommand{\arraystretch}{1.5}
\begin{tabular}{ c | c c } 
\hline
\hline
  $B_{\mathcal{R},J_{z}}$  & $3Q \equiv
                           \frac{\mathcal{N}^{(3)}_{J_{z}}(B_{\mathcal{R}})}{
                           \mathcal{N}^{(3)}_{J_{z}}(B_{\mathcal{R}})
                           +\mathcal{N}^{(5)}_{J_{z}}(B_{\mathcal{R}})}$  
  & $5Q \equiv
    \frac{\mathcal{N}^{(5)}_{J_{z}}(B_{\mathcal{R}})}{
    \mathcal{N}^{(3)}_{J_{z}}(B_{\mathcal{R}}) +
    \mathcal{N}^{(5)}_{J_{z}}(B_{\mathcal{R}})}$  
  \\ 
\hline
$B_{8,1/2}$ & $77\%$ & $23\%$ \\
$B_{10,1/2}$ & $72\%$ & $28\%$ \\
$B_{10,3/2}$ &$74\%$ & $26\%$ \\
\hline
\hline
\end{tabular}
\caption{$3Q$ and $5Q$ fractions of the normalization constants for
  the baryon octet and decuplet in the relativistic
  case. $\mathcal{R}$ stands for the SU(3) representation of the
  baryon multiplet. The spin projection of the corresponding baryon
  multiplet is denoted by $J_{z}$. 
  }
\label{tab:2}
\end{table}
As shown in Tables~\ref{tab:1} and~\ref{tab:2}, the contribution from
the $5Q$ Fock component to the normalization constant is sizable. When
we consider the relativistic case, the fractions of the $5Q$ Fock 
component are reduced by about 10~\%. This indicates that it is
crucial to consider both the $5Q$ Fock component and the relativistic
corrections in the baryon LCWFs. The normalization constants of the
$3Q$ and $5Q$ Fock components have significant physical implications:
The amount of each component directly yields information on how much
the quarks inside a baryon carry the fraction of the baryon
longitudinal momentum. For example, the $3Q$ quarks inside the nucleon
carry 77~\% of the nucleon momentum in the relativistic case.  

\begin{figure}[htp]
\centering
\includegraphics[scale=0.59]{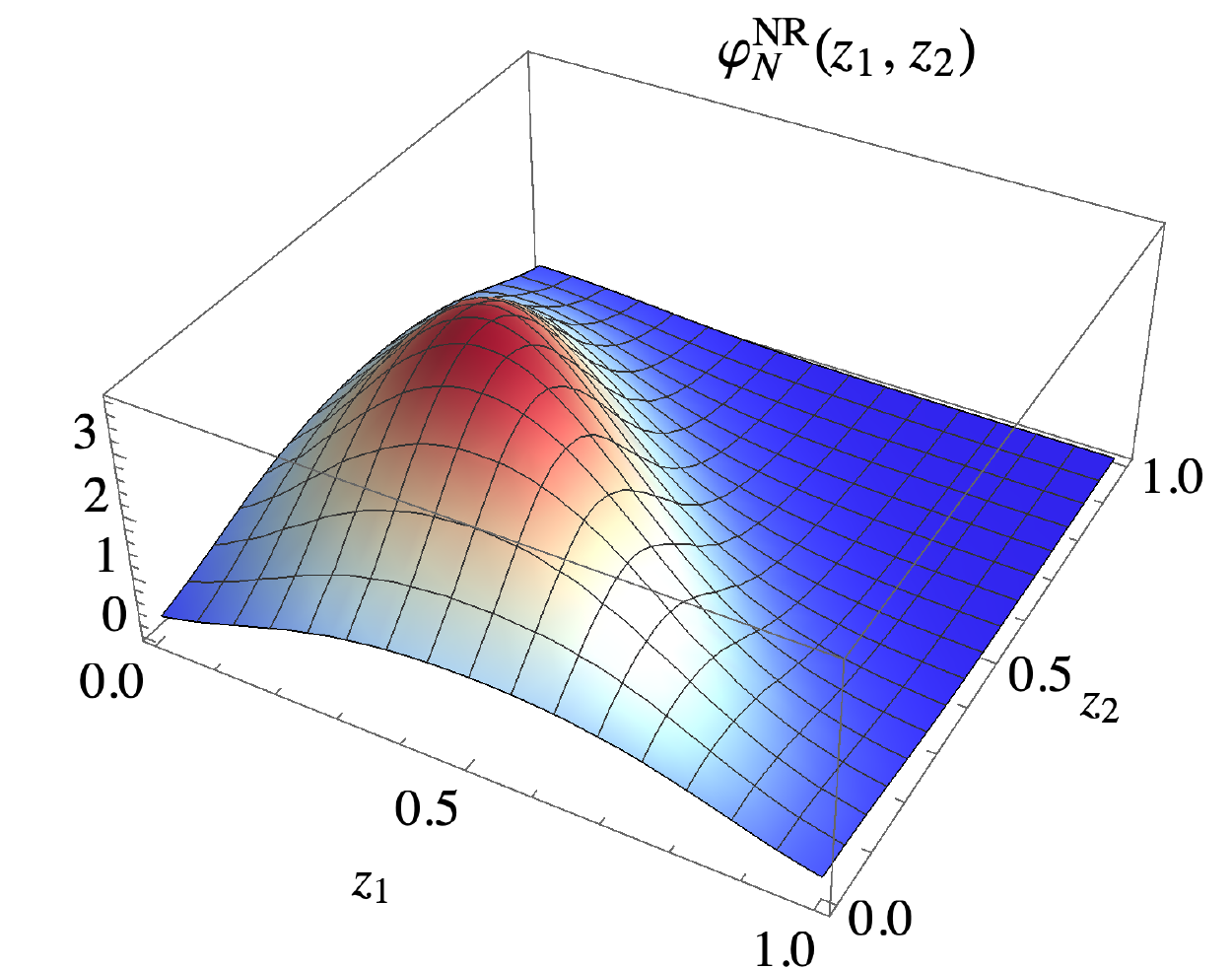}
\includegraphics[scale=0.59]{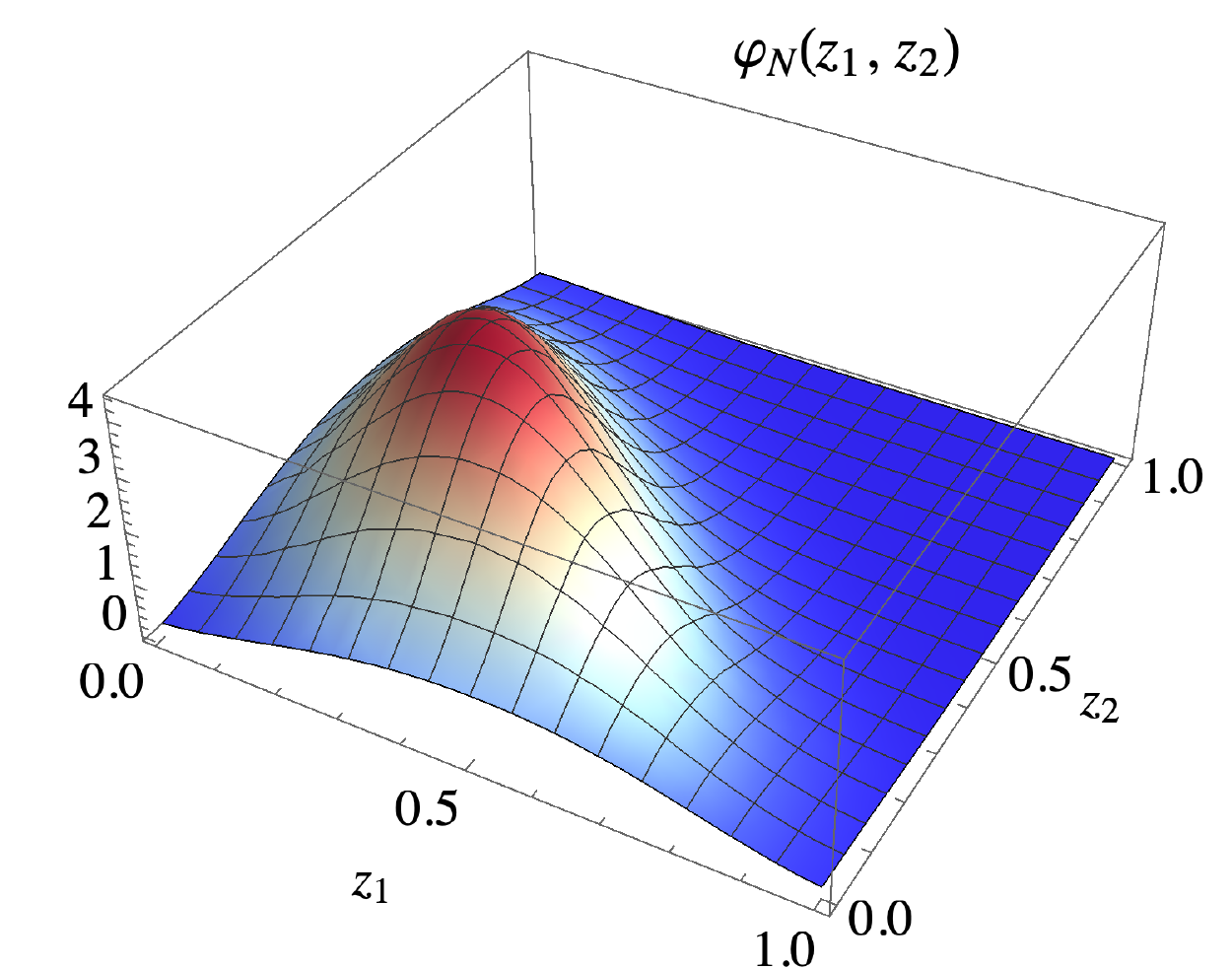}
\caption{Nucleon distribution amplitudes in the non-relativistic (left panel) and relativistic (right panel) cases, given as functions of $z_1$ and $z_2$.}
\label{fig:5}
\end{figure}
In Fig.~\ref{fig:5}, we draw the numerical results for the nucleon DAs  $\varphi_{N}(z_{1},z_{2})\equiv \varphi_{N}(z_{1},z_{2},1-z_{1}-z_{2})$
in both the nonrelativistic (left panel) and relativistic (right panel) cases.
The decomposed results for $F(z_{1},z_{2})$ are presented in Appendix~\ref{appendix:b}.
As shown in Eq.~\eqref{eq:sym_N}, the $\varphi_{N}$ consists of both the
symmetric $V(z_{1},z_{2},z_{3})$ and antisymmetric $A(z_{1},z_{2},z_{3})$ parts
under the exchange $z_{1}\leftrightarrow z_{2}$. From Eq.~\eqref{eq:DAs_Model},
we see that the antisymmetric part of the nucleon DA originates from the relativistic
corrections, which implies that the nucleon DA $\varphi^{\mathrm{NR}}_{N}(z_{1},z_{2})$
becomes symmetric in the nonrelativistic limit when the exchange
$z_{1}\leftrightarrow z_{2}$ is considered, i.e., $A(z_{1},z_{2})=0$.
On the other hand, the antisymmetric DA $A(z_1,z_2)$ comes into
play in the relativistic case, so that the configuration of the nucleon DA
is distorted from the symmetric shape. However, since the antisymmetric DA
contributes only weakly to the nucleon DA (less than 5~\%),
$\varphi_{N}$ turns out almost symmetric. Interestingly, these results are 
rather close to that of the asymptotic nucleon DA, $\varphi^{\mathrm{asy}}_{N} =
120z_{1}z_{2}z_{3}$. The present results are consistent with those from
the lattice QCD~\cite{QCDSF:2008qtn, QCDSF:2008zfe, Gockeler:2008xv, Braun:2014wpa, Bali:2015ykx, RQCD:2019hps}. Note that 
the QCD sum rules~\cite{Chernyak:1984bm, King:1986wi, Chernyak:1987nt, Chernyak:1987nu} give rather asymmetric forms of the nucleon DA. 
 
We want to stress that the present model is valid only in the region
$z N_{c}\sim 1$. This means that the approximation used here 
brings about the non-zero values of the DAs at the endpoints $z=0$ and $z=1$. 
Since at $z=0$ the virtuality of the partons inside a nucleon becomes very large, 
the constant dynamical quark mass is no longer plausible. Thus, it is inevitable to
use the momentum-dependent dynamical quark mass $M(p)$ to explain the
endpoint behavior of the baryon DAs. As pointed out in
Ref.~\cite{Diakonov:1996sr}, the baryon mass is very large in the large $N_c$ limit
($M_B\sim N_c$), so that the recoil effects of the baryon are neglected.
This implies that the baryon DAs may not vanish at $z=1$. 
Moreover, when the partons are very virtual, 
the gluons start to radiate and will affect the DAs considerably,
the Sudakov form factors should be included in the DAs. These 
were neglected here~\cite{Petrov:2002jr, Petrov:1998kf}.
One faces the same problem in the calculation of the structure functions~\cite{Diakonov:1996sr, Wakamatsu:1997en}. 

\begin{figure}[htp]
\centering
\includegraphics[scale=0.59]{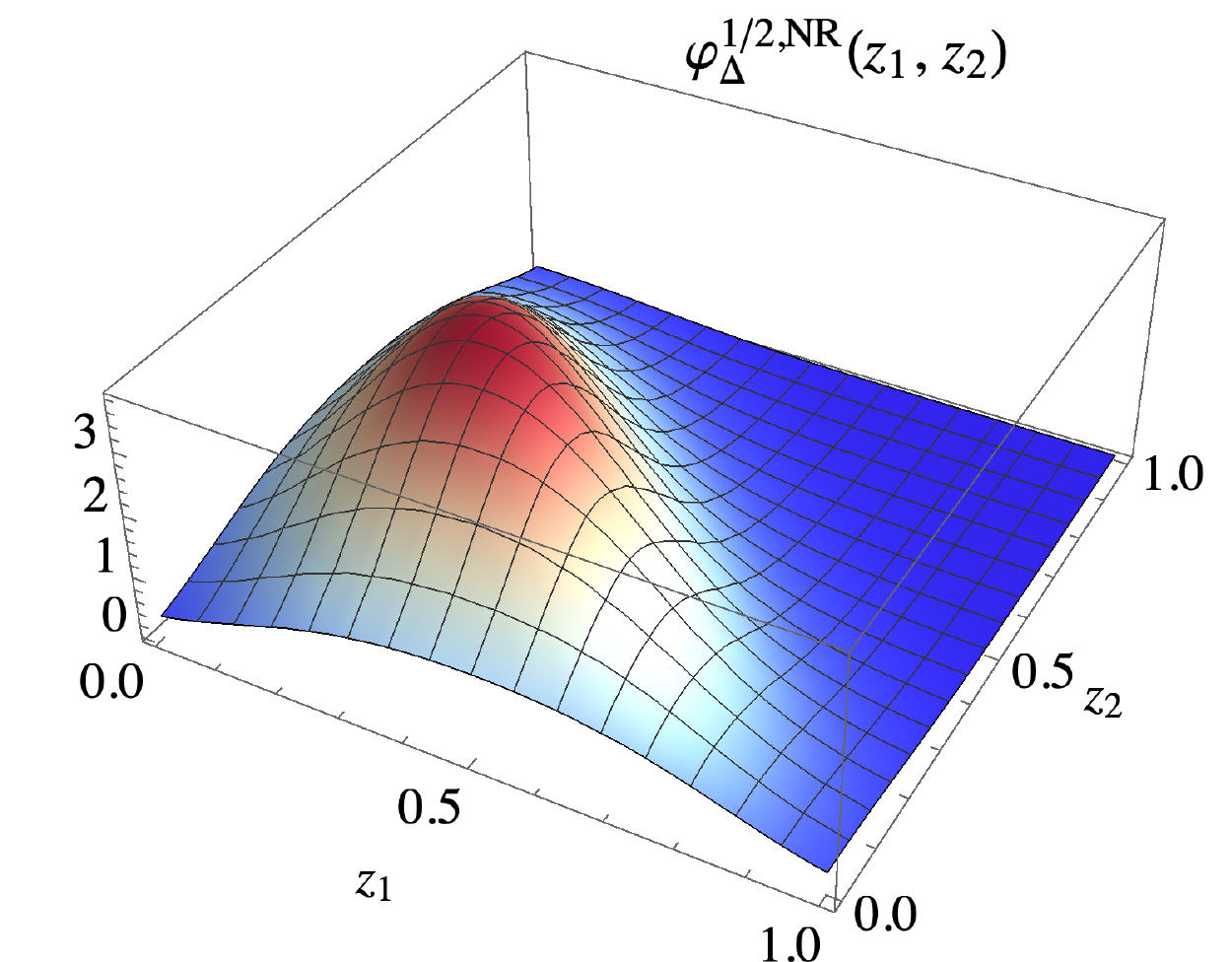}
\includegraphics[scale=0.59]{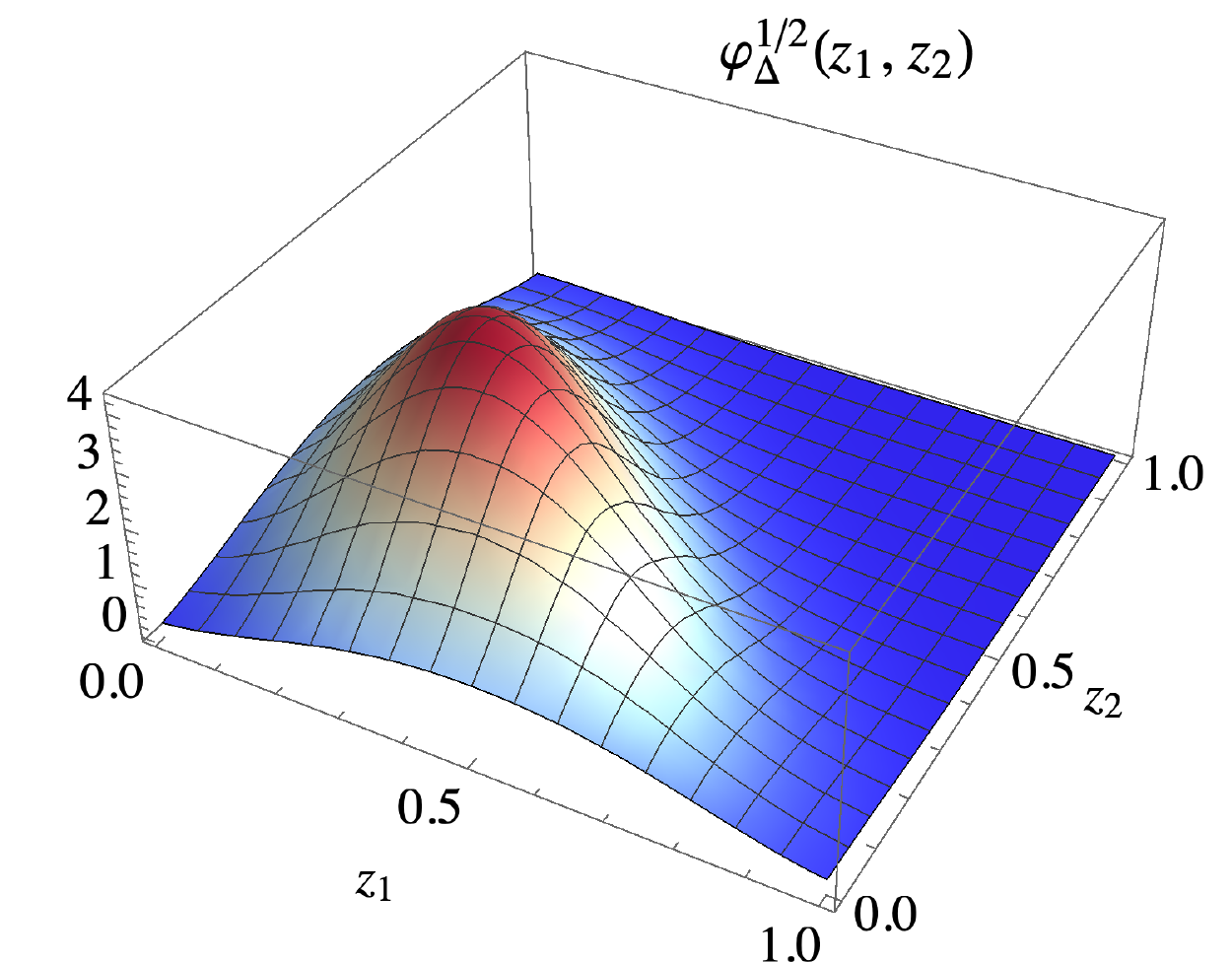}
\includegraphics[scale=0.59]{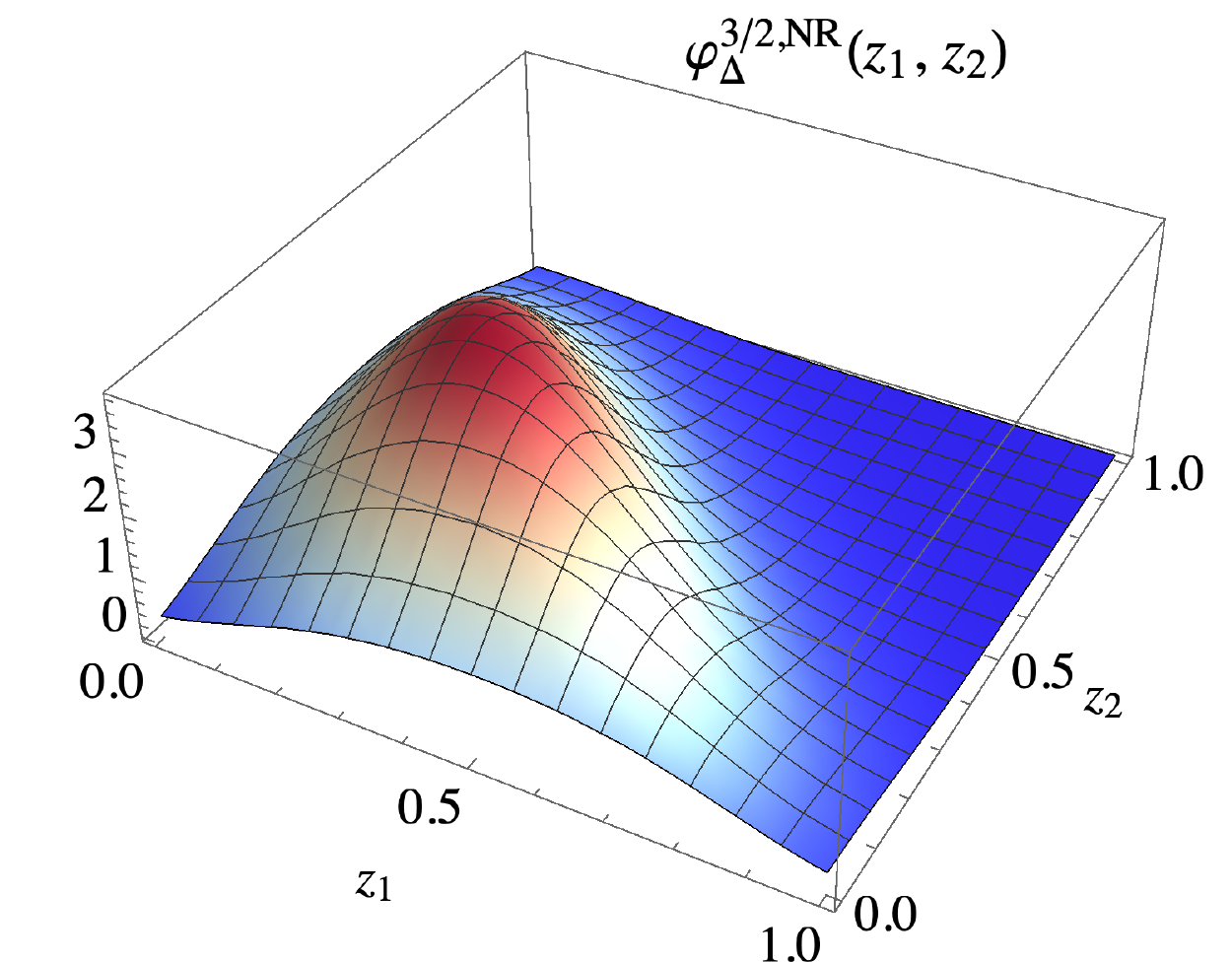}
\includegraphics[scale=0.59]{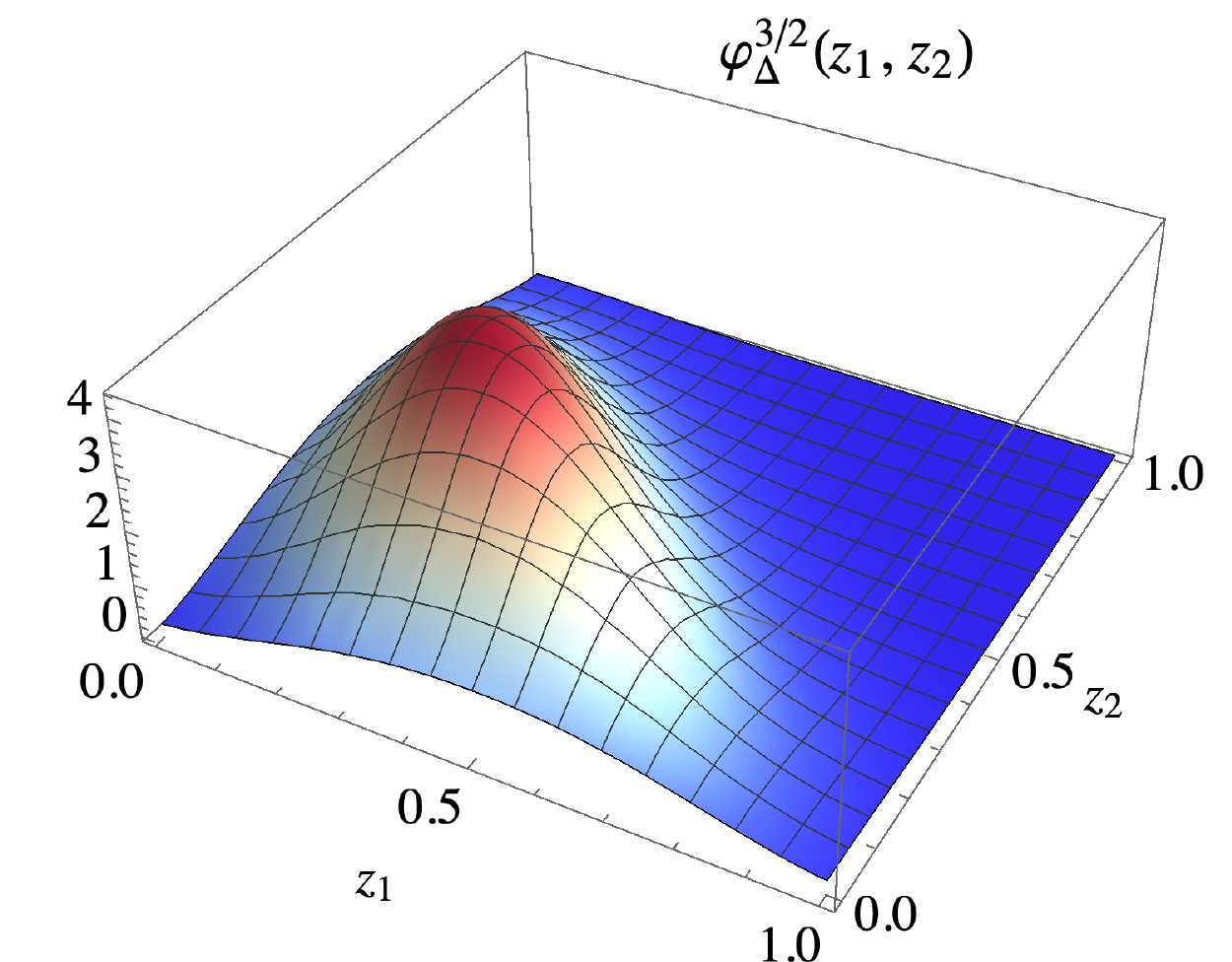}
\caption{$\Delta$ distribution amplitudes in the non-relativistic
  (left panel) and relativistic (right panel) cases.} 
\label{fig:6}
\end{figure}
In Fig.~\ref{fig:6}, we draw the results for the $\Delta$ baryon DAs
with the spin projection 1/2 and 3/2 in non-relativistic and
relativistic cases. The main features are similar to those of the
nucleon shown in Fig.~\ref{fig:5}. The $\Delta$ DAs
$\varphi^{1/2,\mathrm{NR}}_{\Delta},
\varphi^{3/2,\mathrm{NR}}_{\Delta}$ become naturally identical to that
$\varphi^{\mathrm{NR}}_{N}$ of the nucleon. The present results for
the $\Delta$ baryon DAs are also close to the asymptotic DA and are
consistent with those from the QCD sum rules~\cite{Farrar:1988vz}.  

The baryon decay constants or the DA normalization constants, $f_{N},
f^{1/2}_{\Delta}$ and $f^{3/2}_{\Delta}$, contain essential
information on the baryon DAs, which were defined in
Eqs.~\eqref{eq:matrixoctet} and ~\eqref{eq:matrixdecuplet}. We list in
Table~\ref{tab:3} the numerical results for both the $3Q$ and $3Q+5Q$ 
contributions to them in the nonrelativistic and relativistic cases. 
\begin{table}[htp]
\centering
\setlength{\tabcolsep}{5pt}
\renewcommand{\arraystretch}{1.5}
\begin{tabular}{ c | c c c c  } 
\hline
\hline
  & \multicolumn{2}{c}{Non-relativistic} & \multicolumn{2}{c}{Relativistic}    \\
\hline
 $f_{B}$ & $3Q$ & $3Q+5Q $  & $3Q$ & $3Q+5Q $   \\
\hline
$|f_{N}|$ & $6.2$ & $5.0$ & $4.8$ & $4.2$ \\
$|f^{1/2}_{\Delta}|$ & $15.2$ & $11.6$ & $14.9$ & $12.7$ \\
$|f^{3/2}_{\Delta}|$ &$18.7$  &$14.7$   &$15.5$ &$13.4$ \\
\hline
\hline
\end{tabular}
\caption{Non-relativistic and relativisic DA normalizations of the nucleon and $\Delta$
  baryon in units of $10^{-3}~\mathrm{GeV}^{2}$. } 
\label{tab:3}
\end{table}
Interestingly, in the nonrelativistic limit, we have derived the relations
between the DA normalization constants for the nucleon and $\Delta$ baryon
given in Eq.~\eqref{eq:relation_NR}. The numerical results listed in
Table~\ref{tab:3} satisfy these relations. Moreover,
we find that both the $5Q$ component and the relativistic corrections are
significant to determine the DA normalization constants.

These normalization constants are scale-dependent as mentioned already. 
The scale of the present model can be related to the average size of the instanton,
of which the value is $\rho \sim 0.33\,\mathrm{fm}$. So, we will approximately take  
$\mu^{2}_{0}\sim 0.36~\mathrm{GeV}^{2}$ as the scale of the $\chi$QSM~\cite{Kim:1995bq, Diakonov:1995qy, Polyakov:2020cnc}. To compare the present results with those from other works, 
one has to match the scale by using the renormalization group equation for the
DA normalization constants $f_{N},f^{1/2}_{\Delta}$ and $f^{3/2}_{\Delta}$. The
one-loop evolution of them is derived in Refs.~\cite{Lepage:1980fj, Braun:1999te}
\begin{align}
f_{B}(\mu^{2}) = f_{B}(\mu_{0}^{2}) \left(\frac{\alpha_{\mathrm{s}}(\mu^{2})}{\alpha_{\mathrm{s}}(\mu^{2}_{0})}\right)^{\gamma_{0}},
\end{align}
where the anomalous dimension $\gamma_{0} = 2/(3\beta_{0}) $ is given by  Refs.~\cite{Lepage:1980fj, Bergmann:1999ud, Stefanis:1994zd} with the beta function $\beta_{0}=11-2n_{f}/3$ for the strong coupling constant  $\alpha_{\mathrm{s}}(\mu^2)$.  Being evolved to a high normalization point,
for example, $\mu=2~\mathrm{GeV}$, the DA normalization constants are
respectively found to be 
\begin{align}
|f_{N}|=3.9\times 10^{-3}~\mathrm{GeV}^{2}, \ \ |f^{1/2}_{\Delta}|=11.7\times 10^{-3}~\mathrm{GeV}^{2}, \ \ |f^{3/2}_{\Delta}|=12.3\times 10^{-3}~\mathrm{GeV}^{2}.
\end{align}
Though we take $\mu^{2}_{0}\sim 0.36~\mathrm{GeV}^{2}$ as the scale of the
present model, this is not uniquely determined. In fact, it depends on a parameter  varied in the variational estimate of
bulk properties of the instanton medium.  In Refs.~\cite{Kim:1995bq,
  Diakonov:1995qy, Polyakov:2020cnc} the scale of the $\chi$QSM was estimated
to be $\mu_{0}=(0.6-1.4)~\mathrm{GeV}$ with the parameter varied.
Owing to the weak dependence of the DA normalization constants on the
scale evolution, the corresponding error lies within around 5~\%. So, $\mu_0^2 = 0.36\,\mathrm{GeV}^2$ is a reasonable choice. 

In Refs.~\cite{Lorce:2006nq, Lorce:2007as, Lorce:2007xax}, the DA normalization
constant for the nucleon from the $7Q$ contribution $\mathcal{N}^{(7)}$ was
investigated. Though we do not evaluate them explicitly, we use the value
obtained in Refs.~\cite{Lorce:2006nq, Lorce:2007as, Lorce:2007xax} and
estimate $f_{N}$ at $\mu=2~\mathrm{GeV}$:
\begin{align} 
|f_{N}| \simeq 3.7 \times 10^{-3} \mathrm{GeV}^{2} \ \ \  (3Q+5Q+7Q).
\label{eq:fn}
\end{align}
We then expect that further higher Fock contributions are suppressed.
This value was first introduced and estimated in the QCD sum rules~\cite{Chernyak:1984bm, King:1986wi, Chernyak:1987nt, Chernyak:1987nu, Farrar:1988vz}, where $|f_N|$ was estimated to
be $|f_{N}(\mu \simeq 1 \, \mathrm{GeV})| \simeq 5.0 \times 10^{-3} \,
\mathrm{GeV}^{2}$~\cite{Farrar:1988vz}. This value is rather large compared with the present result.
Recently, however, it was obtained from the lattice QCD~\cite{RQCD:2019hps} as follows:   
$|f_{N}(\mu= 2 \, \mathrm{GeV})| = 3.54^{+6}_{-4} \times 10^{-3} \, \mathrm{GeV}^{2}$. 
The present result given in Eq.~\eqref{eq:fn} is in good agreement with this lattice
data. For the $\Delta$ baryon, the QCD sum rules~\cite{Farrar:1988vz}  yield the
following results: $|f^{1/2}_{\Delta}(\mu\simeq 1 \, \mathrm{GeV})|=
(12 \pm 0.2) \times 10^{-3} \, \mathrm{GeV}^{2}$ and
$ |f^{3/2}_{\Delta}(\mu\simeq1 \, \mathrm{GeV})|= 14  \times 10^{-3} \,
\mathrm{GeV}^{2}$, which are consistent with the present ones. 

\section{Summary and conclusions \label{sec:6}}
In the present work, we aimed at providing the leading-twist distribution
amplitudes of the nucleon and $\Delta$ baryon and their normalization constants
$f_{N}, f^{1/2}_{\Delta}$ and $f^{3/2}_{\Delta}$ within the chiral
quark-soliton model. 
We  first defined the distribution amplitudes by means of the
vacuum-to-baryon matrix elements of trilocal QCD operators. The leading-twist
distribution amplitudes are related to the light-cone wave functions with the orbital
angular momentum $L_{z}=0$ involved in the valence three-quark Fock component.
In the chiral quark-soliton model, the explicit light-cone wave
functions of the baryon octet and decuplet are derived,
which consist of the valence quark and quark-antiquark
(pair) wave functions. Expanding the vacuum wave function or the Dirac sea,
we were able to construct the light-cone wave function for the higher
Fock components, i.e, 3, 5, 7...-wave functions. As pointed out in
Refs.~\cite{Diakonov:2005ib, Lorce:2007as}, it turned out that the five-quark
contributions significantly contribute to the normalization for the baryon
light-cone wave functions and are found to be $\sim 15\%$.
Having included the five-quark contributions, we estimated the distribution
amplitudes of the nucleon and $\Delta$ baryon.
The nucleon distribution amplitude $\varphi_{N}$ is found to be almost
symmetric under the exchange of the first two arguments,
which is consistent with the recent results from the
lattice QCD~\cite{RQCD:2019hps, Braun:2014wpa}. Interestingly,
the normalization constants for the proton distribution amplitude
are obtained to be $|f_{N}(\mu=2 \, \mathrm{GeV})|= 3.7 \times 10^{-3}
\, \mathrm{GeV}^{2}$, which is very close to the lattice data,
$f_{N}(\mu\sim2 \, \mathrm{GeV})= 3.54 \times 10^{-3} \,
\mathrm{GeV}^{2}$~\cite{RQCD:2019hps}. The distribution amplitudes of
the $\Delta$ baryon $\varphi^{1/2}_{\Delta}$ and $\varphi^{3/2}_{\Delta}$ are
respectively found to be almost and totally symmetric, and their
normalization constants are respectively obtained to be $|f^{1/2}_{\Delta}(\mu=2 \,
\mathrm{GeV})|= 11.7 \times 10^{-3} \, \mathrm{GeV}^{2 }$ and
$ |f^{3/2}_{\Delta}(\mu=2 \, \mathrm{GeV})|= 12.3 \times 10^{-3} \,
\mathrm{GeV}^{2}$.
We anticipate that the results from the lattice QCD or several theoretical
approaches for the $\Delta$ baryon distribution amplitudes will soon
come out.

The nucleon and $\Delta$ baryon light-cone wave functions obtained from the
present work can be applied to the study of the electromagnetic and mechanical
properties of the nucleon and $\Delta$. Since the
electromagnetic and gravitational form factors can be
directly derived from the light-cone wave functions, we can immediately get access to
the transverse charge densities for the nucleon and $\Delta$ baryon.
More importantly, the mechanical properties of both the nucleon and $\Delta$
with the stability conditions can be scrutinized by using the present results,
since they provide essential information on how the
nucleon~\cite{Polyakov:2018zvc,Kim:2021jjf, Alharazin:2020yjv,
  Panteleeva:2021iip, Gegelia:2021wnj} and
$\Delta$~\cite{Polyakov:2018rew, Panteleeva:2020ejw, Kim:2020lrs} are
shaped physically. On the other hand, it is also of great importance
to investigate the baryon light-cone wave functions with the momentum-dependent
dynamical quark mass obtained from the instanton vacuum.
This will make the baryon light-cone wave functions to satisfy the correct endpoint
behavior. Recently, the chiral quark-soliton model or the pion mean-field approach
has been successfully extended to the description of singly heavy
baryons~\cite{Yang:2016qdz, Kim:2017jpx, Kim:2017khv, Kim:2018xlc,
  Yang:2018uoj, Kim:2018nqf, Kim:2018cxv, Kim:2019rcx, Yang:2019tst,
  Kim:2019wbg, Yang:2020klp,  Kim:2020nug, Kim:2020uqo}.
Thus, it is of great interest to study the light-cone wave functions
of the singly heavy baryons. The corresponding studies are under way. 

\begin{acknowledgments}
The present work was supported by Basic Science Research Program
through the National Research Foundation of Korea funded by the Korean
government (Ministry of Education, Science and Technology, MEST),
Grant-No. 2021R1A2C2093368 and 2018R1A5A1025563.
J.-Y.K is supported by the Deutscher Akademischer
Austauschdienst(DAAD) doctoral  scholarship. 
\end{acknowledgments}

\clearpage
\appendix

\section{Baryon rotational wave functions}
\label{appendix:a}
 The wave functions for the baryon octet are generically expressed as
mixed tensors, $P_f^g$, whereas those for the  baryon decuplet as
symmetric ones, $D_{f_1f_2f_3}$ with three quark indices.
This means that the Wigner $D$ functions for the baryon octet and
decuplet can be expressed as follows:
\begin{align}
&[D^{(8,1/2)*}(R)]^{g}_{f,k} \sim \epsilon_{kl} R^{\dagger l}_{f} R^{g}_{3}, \cr
&[D^{(10,3/2)*}(R)]^{g}_{\{f_1f_2f_3\},\{k_1k_2k_3\}} \sim \epsilon_{k'_{1}k_{1}}\epsilon_{k'_{2}k_{2}}\epsilon_{k'_{3}k_{3}} R^{\dagger k'_1}_{f_1}R^{\dagger k'_2}_{f_2}R^{\dagger k'_3}_{f_3} |_{\{f_1f_2f_3\}},
\end{align}
where $\epsilon_{kl}$ stands for the antisymmetric tensor. The $k=1$
means the spin-up baryon state, whereas $k=2$ represents the spin-down
state. The wave functions for the baryon decuplet are fully
symmetrized in the flavor $\{f_1f_2f_3\}$ and spin projection
$\{k_1k_2k_3\}$ indices. The flavor part of the baryon octet are
explicitly written as~\cite{Fayyazuddin:2012qfa}:
\begin{align}
&P^{3}_{1} = N^{+}, \ \ \ P^{3}_{2} =
                N^{0}, \ \  \ P^{2}_{1}=\Sigma^{+}, \ \ 
                \ P^{1}_{2}=\Sigma^{-}, \cr 
&P^{1}_{1}=\frac{1}{\sqrt{2}}\Sigma^{0} +
\frac{1}{\sqrt{6}}\Lambda^{0}, \ \  \
  P^{2}_{2}=-\frac{1}{\sqrt{2}}\Sigma^{0} +
                                               \frac{1}{\sqrt{6}}\Lambda^{0},
                                               \cr   
  &P^{3}_{3} = -\sqrt{\frac{2}{3}}\Lambda^{0}, \ \ \
    P^{2}_{3} =  \Xi^{0}, \ \ \
 P^{1}_{3} =-\Xi^{-},
\end{align}
and that of the decuplet $D_{\{f_1f_2f_3\}}$ are given as
\begin{align}
  &D_{111} =\sqrt{6} \Delta^{++}, \ \ \
    D_{112} =\sqrt{2} \Delta^{+}, \ \  \
    D_{122} =\sqrt{2} \Delta^{0}, \ \  \
    D_{222} =\sqrt{6} \Delta^{-}, \cr
  &D_{113}=\sqrt{2}\Sigma^{*+}, \ \  \
    D_{123}=-\Sigma^{*0}, \ \ \
    D_{223}=-\sqrt{2}\Sigma^{*-}, \ \ \
    D_{133}=\sqrt{2}\Xi^{*0}, \cr
  &D_{233}=\sqrt{2}\Xi^{*-}, \ \ \
    D_{333}=-\sqrt{6}\Omega^{-}.
\end{align}
The prefactors for the wave functions are determined by normalizing the
rotational wave functions 
\begin{align}
\int dR B^{*}_{\mathrm{spin}}(R)B^{\mathrm{spin}}(R)=1.
\end{align}
For example, the nucleon state is explcitly expressed as 
\begin{align}
 N^{+*}_{k}(R)= \sqrt{8} \epsilon_{kl} R^{\dagger l}_{1}R^{3}_{3}, \ \ \
  N^{0*}_{k}(R)= \sqrt{8} \epsilon_{kl} R^{\dagger l}_{2}R^{3}_{3},
\end{align}
whereas the $\Delta$ baryon with spin projection 3/2 ($\uparrow
\uparrow \uparrow$) and 1/2~($\uparrow$) are written as 
\begin{align}
  &\Delta^{++*}_{\uparrow\uparrow\uparrow}(R)=
    \sqrt{10} R^{\dagger2}_{1}R^{\dagger2}_{1}R^{\dagger2}_{1}, \ \ \
    \Delta^{+*}_{\uparrow\uparrow\uparrow}(R)=
    \sqrt{30} R^{\dagger2}_{1}R^{\dagger2}_{1}R^{\dagger2}_{2}, \cr
&\Delta^{++*}_{\uparrow}(R)= \sqrt{30}
R^{\dagger2}_{1}R^{\dagger2}_{1}
R^{\dagger1}_{1}, \ \ \ \Delta^{+*}_{\uparrow}(R)= \sqrt{10}
 (R^{\dagger2}_{1}R^{\dagger2}_{1}R^{\dagger1}_{2} + 2R^{\dagger2}_{2}
  R^{\dagger2}_{1}R^{\dagger1}_{1}). 
\end{align}
For different spin projections, the integral becomes zero. The
rotational wave functions belonging to different baryons satisfy the
orthogonality.

\section{Distribution amplitudes }
\label{appendix:b}
The distribution amplitudes are listed for the nucleon
\begin{align}
  &f_{N}T_{N} = -12\sqrt{6} c^{\bm{8},1/2}_{0} \int [dk_{\perp}]
    T^{112}_{j_{1}j_{2}j_{3},1} F^{j_{1}1}(\bm{p}_{1})
    F^{j_{2}1}(\bm{p}_{2})F^{j_{3}2}(\bm{p}_{3}), \cr
  &f_{N}V_{N} = 12\sqrt{6} c^{\bm{8},1/2}_{0} \int [dk_{\perp}]
    T^{112}_{j_{1}j_{2}j_{3},1} \bigg{[}F^{j_{1}2}(\bm{p}_{1})
    F^{j_{2}1}(\bm{p}_{2})F^{j_{3}1}(\bm{p}_{3})  \cr
  & \hspace{5.6cm}+ F^{j_{1}1}(\bm{p}_{1})F^{j_{2}2}(\bm{p}_{2})
    F^{j_{3}1}(\bm{p}_{3})\bigg{]}, \cr
  &f_{N}A_{N} = 12\sqrt{6} c^{\bm{8},1/2}_{0} \int [dk_{\perp}]
    T^{112}_{j_{1}j_{2}j_{3},1} \bigg{[}F^{j_{1}2}(\bm{p}_{1})
    F^{j_{2}1}(\bm{p}_{2})F^{j_{3}1}(\bm{p}_{3})  \cr
  & \hspace{5.6cm}- F^{j_{1}1}(\bm{p}_{1})
    F^{j_{2}2}(\bm{p}_{2})F^{j_{3}1}(\bm{p}_{3})\bigg{]}, 
\end{align}
and for the $\Delta$ baryon
\begin{align}
  &f^{1/2}_{\Delta}T_{\Delta} = -12\sqrt{6} c^{\bm{10},1/2}_{0}
    \int [dk_{\perp}] T^{111}_{j_{1}j_{2}j_{3},1}
    F^{j_{1}1}(\bm{p}_{1}) F^{j_{2}1}(\bm{p}_{2})F^{j_{3}2}(\bm{p}_{3}), \cr
  &f^{1/2}_{\Delta}V_{\Delta} = -12\sqrt{6} c^{\bm{10},1/2}_{0}
    \int [dk_{\perp}] T^{111}_{j_{1}j_{2}j_{3},1} \bigg{[}
    F^{j_{1}2}(\bm{p}_{1})F^{j_{2}1}(\bm{p}_{2})F^{j_{3}1}(\bm{p}_{3})  \cr
  & \hspace{6.3cm} + F^{j_{1}1}(\bm{p}_{1})F^{j_{2}2}(\bm{p}_{2})
    F^{j_{3}1}(\bm{p}_{3})\bigg{]}, \cr
  &f^{1/2}_{\Delta}A_{\Delta} = -12\sqrt{6} c^{\bm{10},1/2}_{0}
    \int [dk_{\perp}] T^{111}_{j_{1}j_{2}j_{3},1}
    \bigg{[}F^{j_{1}2}(\bm{p}_{1}) F^{j_{2}1}(\bm{p}_{2})F^{j_{3}1}(\bm{p}_{3})  \cr
  & \hspace{6.3cm} - F^{j_{1}1}(\bm{p}_{1})F^{j_{2}2}(\bm{p}_{2})
    F^{j_{3}1}(\bm{p}_{3})\bigg{]}, 
\end{align}
By summing over the isospin indices, we obtain the explicit
expressions of the distribution amplitudes for the nucleon  
\begin{align}
f_{N}T_{N} &= 2\sqrt{3} c^{\bm{8},1/2}_{0} \int [dk_{\perp}]
             \bigg{[}2f_{\parallel}(\bm{\bm{k}_{1}})f_{\parallel}(\bm{\bm{k}_{2}})
             f_{\parallel}(\bm{\bm{k}_{3}}) 
             \cr 
& +k_{1L}k_{3R}f_{\perp}(\bm{\bm{k}_{1}})f_{\perp}(\bm{\bm{k}_{3}})f_{\parallel}(\bm{\bm{k}_{2}})+k_{2L}k_{3R}f_{\perp}(\bm{\bm{k}_{2}})f_{\perp}(\bm{\bm{k}_{3}})f_{\parallel}(\bm{\bm{k}_{1}})\bigg{]}, \cr
f_{N}V_{N} &= 2\sqrt{3} c^{\bm{8},1/2}_{0} \int [dk_{\perp}] \bigg{[}2f_{\parallel}(\bm{\bm{k}_{1}})f_{\parallel}(\bm{\bm{k}_{2}})f_{\parallel}(\bm{\bm{k}_{3}})   \cr
& -k_{2L}k_{1R}f_{\perp}(\bm{\bm{k}_{2}})f_{\perp}(\bm{\bm{k}_{1}})f_{\parallel}(\bm{\bm{k}_{3}})+k_{1L}k_{2R}f_{\perp}(\bm{\bm{k}_{1}})f_{\perp}(\bm{\bm{k}_{2}})f_{\parallel}(\bm{\bm{k}_{3}}) \cr
&+2k_{3L}k_{1R}f_{\perp}(\bm{\bm{k}_{3}})f_{\perp}(\bm{\bm{k}_{1}})f_{\parallel}(\bm{\bm{k}_{2}})+2k_{3L}k_{2R}f_{\perp}(\bm{\bm{k}_{3}})f_{\perp}(\bm{\bm{k}_{2}})f_{\parallel}(\bm{\bm{k}_{1}})\bigg{]}, \cr
f_{N}A_{N} &= 2\sqrt{3} c^{\bm{8},1/2}_{0} \int [dk_{\perp}] \bigg{[}-k_{2L}k_{1R}f_{\perp}(\bm{\bm{k}_{2}})f_{\perp}(\bm{\bm{k}_{1}})f_{\parallel}(\bm{\bm{k}_{3}}) \cr
&+k_{1L}k_{2R}f_{\perp}(\bm{\bm{k}_{1}})f_{\perp}(\bm{\bm{k}_{2}})f_{\parallel}(\bm{\bm{k}_{3}})   +2k_{3L}k_{1R}f_{\perp}(\bm{\bm{k}_{3}})f_{\perp}(\bm{\bm{k}_{1}})f_{\parallel}(\bm{\bm{k}_{2}}) \cr
&-2k_{3L}k_{2R}f_{\perp}(\bm{\bm{k}_{3}})f_{\perp}(\bm{\bm{k}_{2}})f_{\parallel}(\bm{\bm{k}_{1}})\bigg{]},
\end{align}
and for the $\Delta$ baryon
\begin{align}
f^{1/2}_{\Delta}T_{\Delta}&= -\frac{12}{\sqrt{5}} c^{\bm{10},1/2}_{0} \int [dk_{\perp}] \bigg{[}f_{\parallel}(\bm{k}_{1})f_{\parallel}(\bm{k}_{2})f_{\parallel}(\bm{k}_{3}) \cr
&+k_{1L}k_{3R}f_{\perp}(\bm{k}_{1})f_{\perp}(\bm{k}_{3})f_{\parallel}(\bm{k}_{2})+k_{2L}k_{3R}f_{\perp}(\bm{k}_{2})f_{\perp}(\bm{k}_{3})f_{\parallel}(\bm{k}_{1})\bigg{]}, \cr
f^{1/2}_{\Delta}V_{\Delta} &= -\frac{12}{\sqrt{5}} c^{\bm{10},1/2}_{0} \int [dk_{\perp}] \bigg{[}2f_{\parallel}(\bm{k}_{1})f_{\parallel}(\bm{k}_{2})f_{\parallel}(\bm{k}_{3})  \cr
&-k_{2L}k_{1R}f_{\perp}(\bm{k}_{2})f_{\perp}(\bm{k}_{1})f_{\parallel}(\bm{k}_{3})-k_{1L}k_{2R}f_{\perp}(\bm{k}_{1})f_{\perp}(\bm{k}_{2})f_{\parallel}(\bm{k}_{3}) \cr
&-k_{3L}k_{1R}f_{\perp}(\bm{k}_{3})f_{\perp}(\bm{k}_{1})f_{\parallel}(\bm{k}_{2})-k_{3L}k_{2R}f_{\perp}(\bm{k}_{3})f_{\perp}(\bm{k}_{2})f_{\parallel}(\bm{k}_{1})\bigg{]}, \cr
f^{1/2}_{\Delta}A_{\Delta} &= \frac{12}{\sqrt{5}} c^{\bm{10},1/2}_{0} \int [dk_{\perp}] \bigg{[}k_{2L}k_{1R}f_{\perp}(\bm{k}_{2})f_{\perp}(\bm{k}_{1})f_{\parallel}(\bm{k}_{3}) \cr
&-k_{1L}k_{2R}f_{\perp}(\bm{k}_{1})f_{\perp}(\bm{k}_{2})f_{\parallel}(\bm{k}_{3}) +k_{3L}k_{1R}f_{\perp}(\bm{k}_{3})f_{\perp}(\bm{k}_{1})f_{\parallel}(\bm{k}_{2}) \cr
&-k_{3L}k_{2R}f_{\perp}(\bm{k}_{3})f_{\perp}(\bm{k}_{2})f_{\parallel}(\bm{k}_{1})\bigg{]}.
\end{align}

\begin{figure}[htp]
\centering
\includegraphics[scale=0.59]{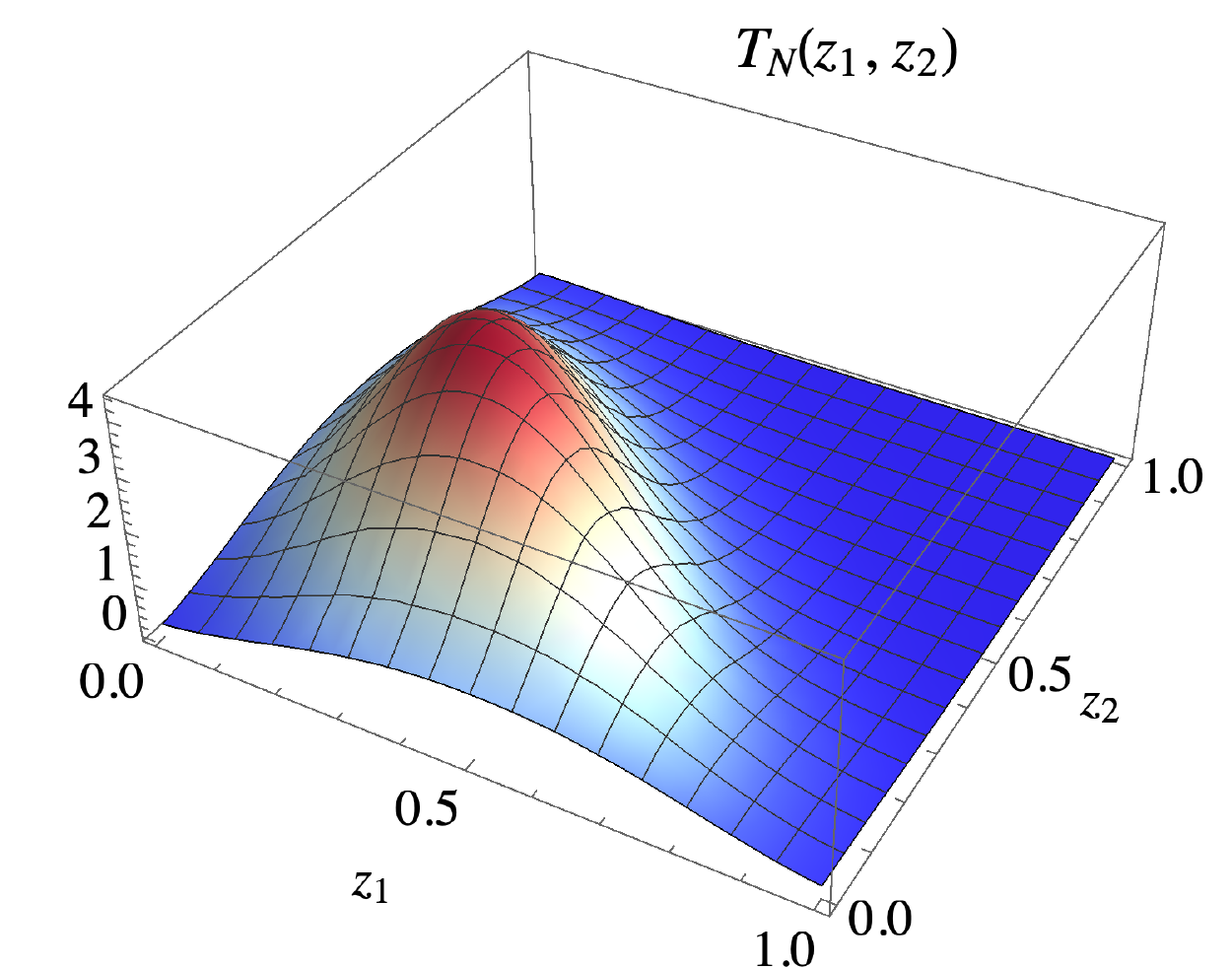}
\includegraphics[scale=0.59]{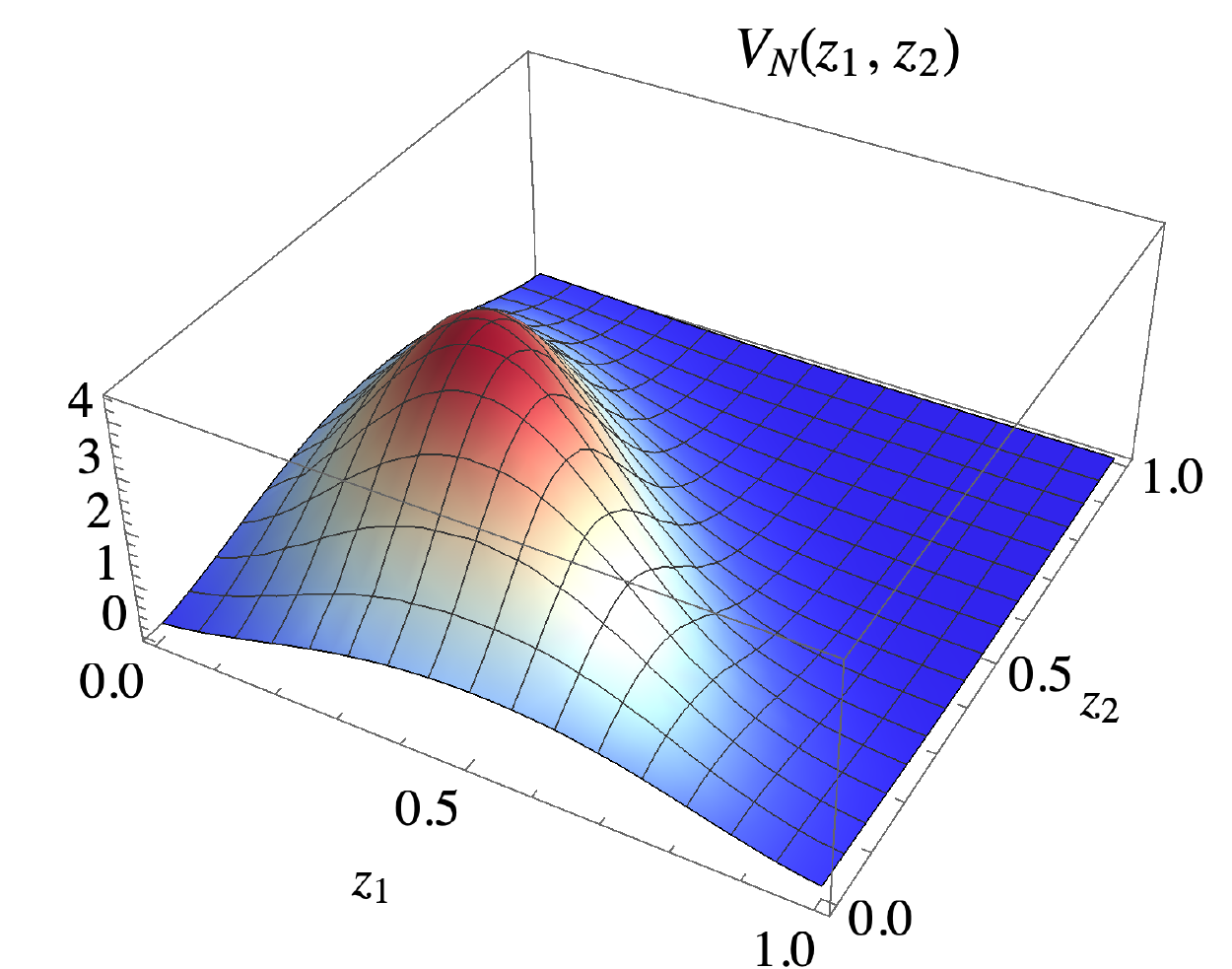}
\includegraphics[scale=0.59]{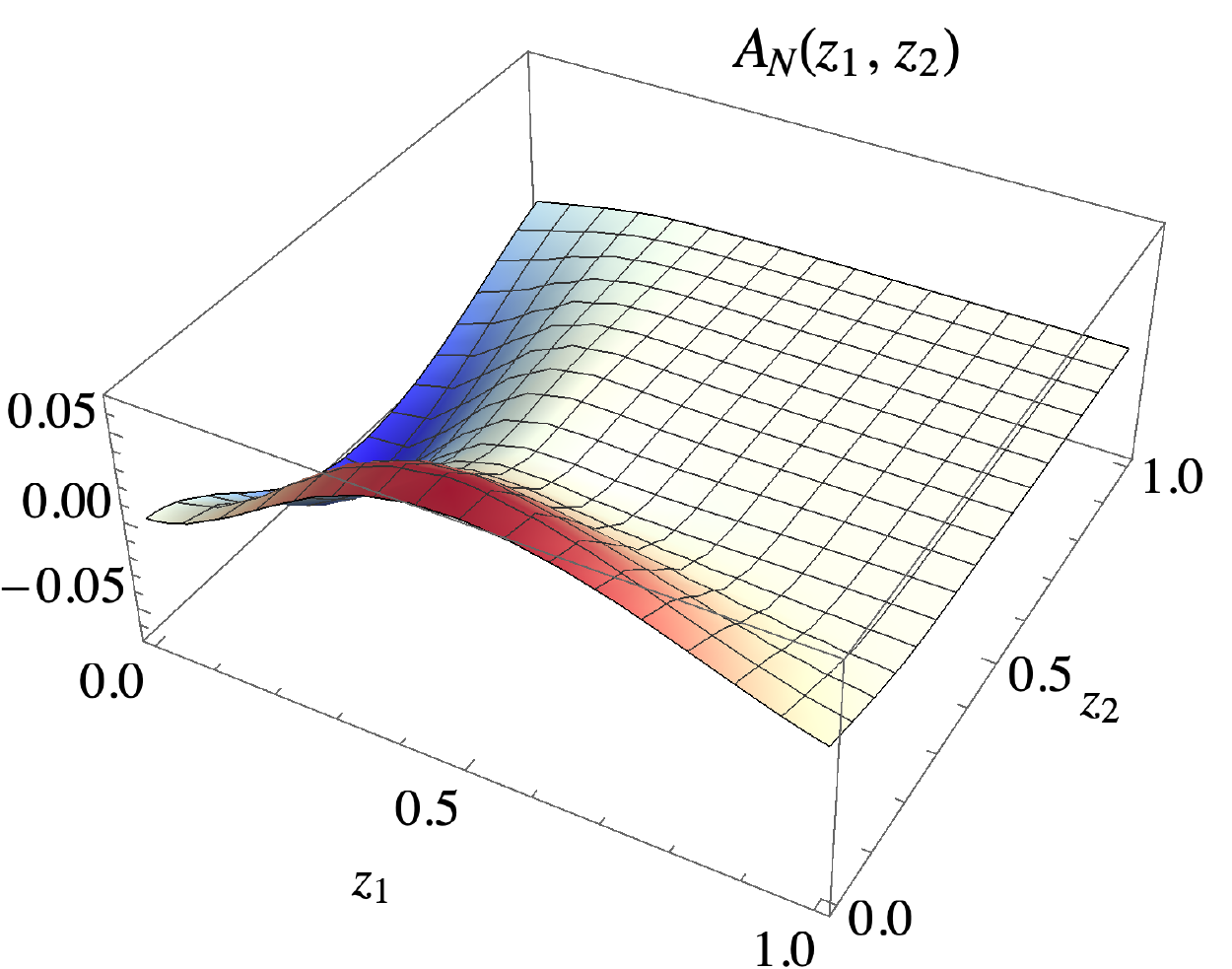}
\caption{Nucleon distribution amplitudes $T_{N},V_{N}$ and $A_{N}$}
\label{fig:7}
\end{figure}
In Fig.~\ref{fig:7}, the decomposed distribution amplitudes of the
nucleon are shown. The antisymmetric distribution amplitude $A_{N}$ is
numerically rather small compared with the symmetric ones $V_{N}$ and
$T_{N}$. 

\begin{figure}[htp]
\centering
\includegraphics[scale=0.59]{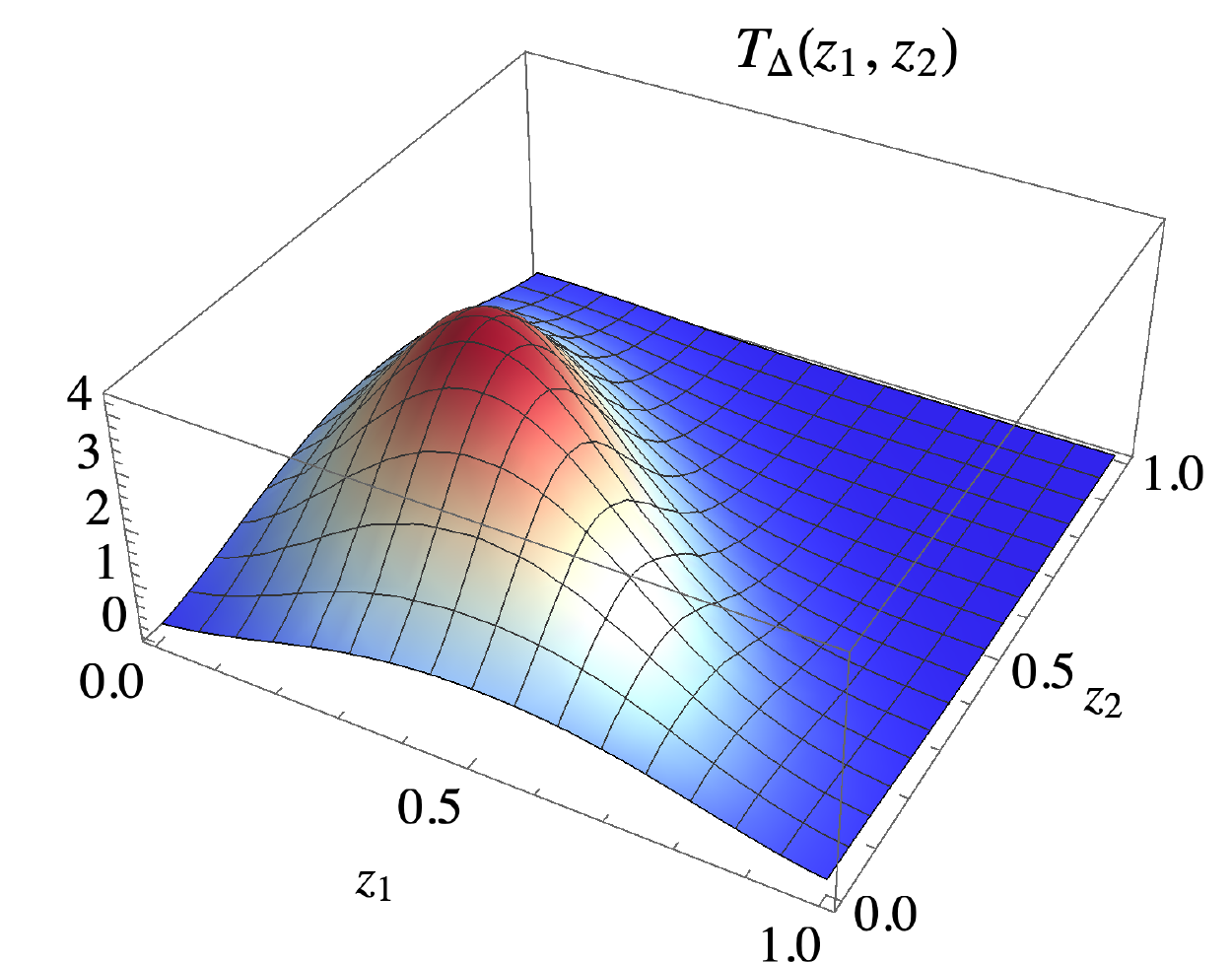}
\includegraphics[scale=0.59]{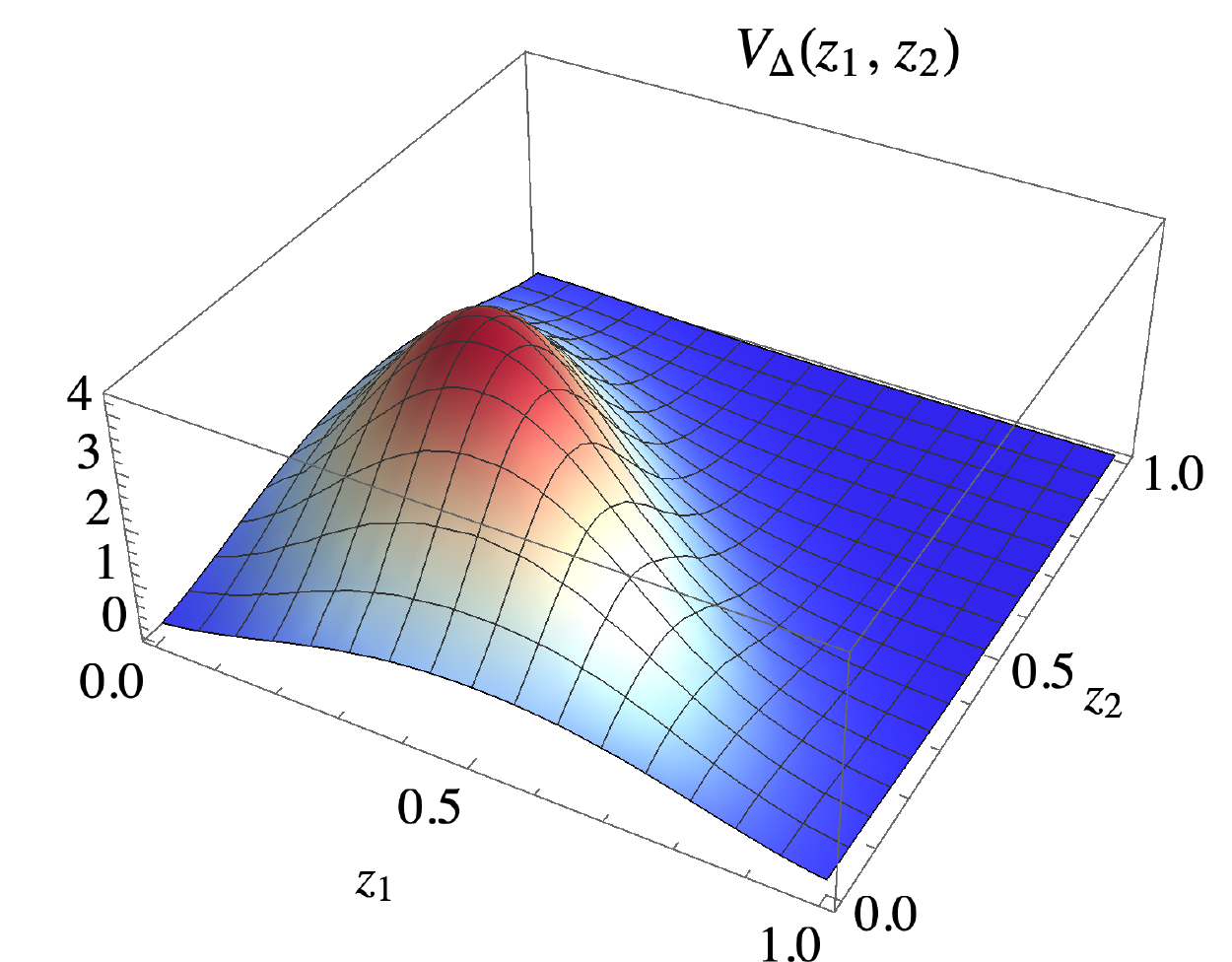}
\includegraphics[scale=0.59]{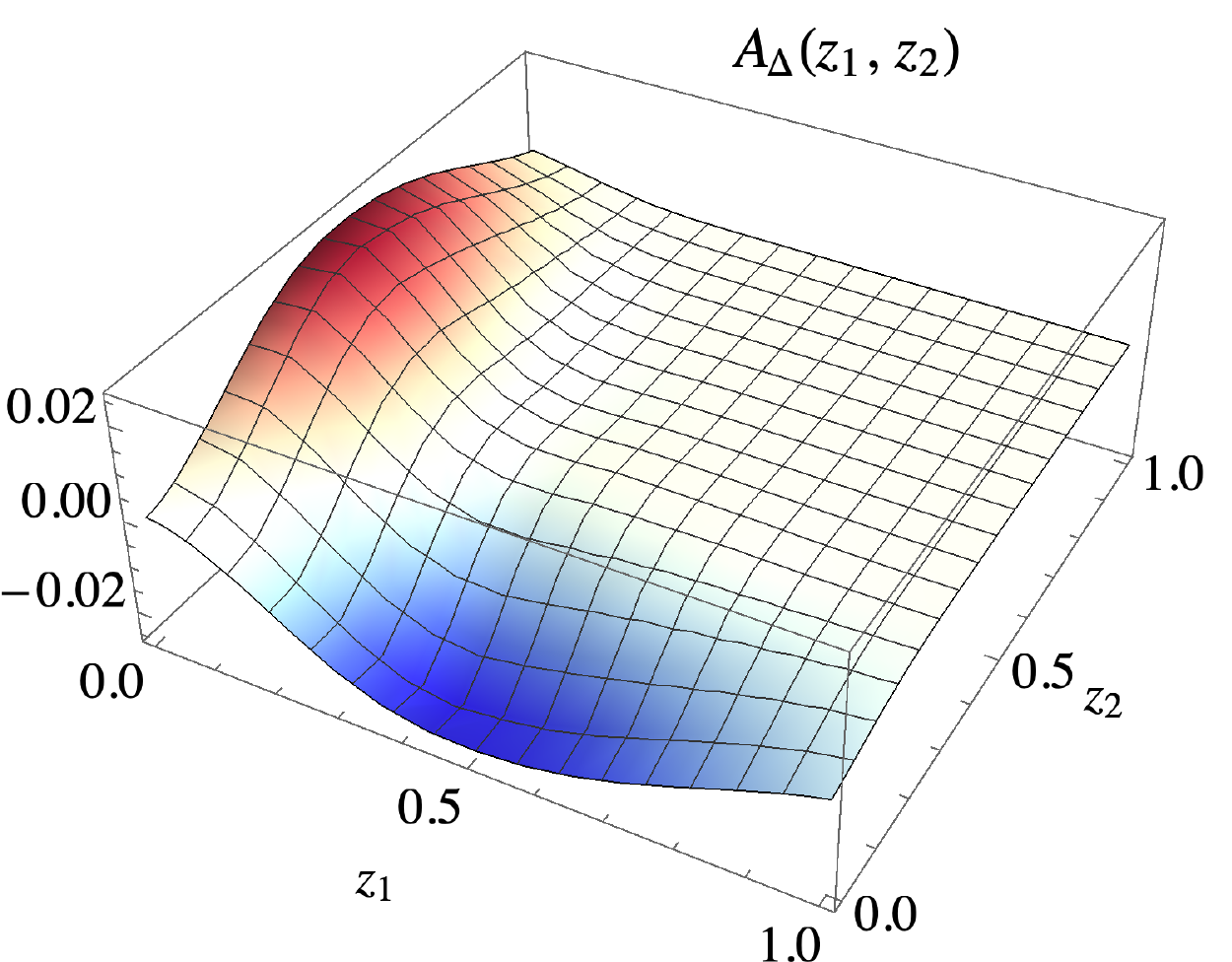}
\caption{$\Delta$ baryon distribution amplitudes
  $T_{\Delta},V_{\Delta}$ and $A_{\Delta}$} 
\label{fig:8}
\end{figure}
In Fig.~\ref{fig:8}, the decomposed distribution amplitudes of the
$\Delta$ baryon are shown. The antisymmetric distribution amplitude
$A_{\Delta}$ is numerically rather small compared with the symmetric
ones $V_{\Delta}$ and $T_{\Delta}$.

\newpage
\bibliographystyle{JHEP}
\bibliography{Nucleon_DAs_20211020_Resubmitted_Final.bib}

\end{document}